\RequirePackage{fix-cm}
\documentclass[twocolumn]{svjour3}          % twocolumn

\usepackage{cite}
\usepackage{amsmath,amssymb,amsfonts}
\usepackage{algorithmic}
\usepackage{graphicx}
\usepackage{float}  % provide floating env
\usepackage{subfigure}

\usepackage{authblk}

\usepackage{textcomp}
\usepackage[flushleft]{caption}
\usepackage{verbatim}
\usepackage{enumerate}
\usepackage{stfloats}
\usepackage{booktabs}% provide \toprule、\midrule、\bottomrule
\usepackage{multirow}% provide  \multirow{}{}{}
\usepackage{hyperref}
\hypersetup{hidelinks,
	colorlinks=true,
	allcolors=black,
	pdfstartview=Fit,
	breaklinks=true}
	
\usepackage{ulem}	

\usepackage{soul, color, xcolor}
\sethlcolor{green}
\soulregister{\cite}7 % 注册\cite命令
\soulregister{\citep}7 % 注册\citep命令
\soulregister{\citet}7 % 注册\citet命令
\soulregister{\ref}7  % 注册\ref命令
\soulregister{\pageref}7 %注册\pageref命令
\soulregister{\eqref} 7
\soulregister{\textcomp}7
\soulregister{\amsmath}7
\soulregister{\amssymb}7
\soulregister{\amsfonts}7
\soulregister{\graphicx}7
\soulregister{\algorihmic}7

\usepackage{bbding}

\smartqed  % flush right qed marks, e.g. at end of proof
\usepackage{graphicx}
\title{CycleGuardian: A Framework for Automatic Respiratory  Sound  classification  Based  on Improved  Deep  clustering  and Contrastive Learning}

\author{Yun Chu         \and 
        Qiuhao Wang     \and 
        Enze Zhou      \and 
        Ling Fu        \and 
        Qian Liu       \and
        Gang Zheng 
         }

\begin{document}

% \title{CycleGuardian: A Framework for Automatic Respiratory  Sound  classification  Based  on Improved  Deep  clustering  and Contrastive Learning}

\titlerunning{CycleGuardian}        % if too long for running head

% \author{Yun Chu         \and 
%         Qiuhao Wang     \and 
%         Enze Zhou      \and 
%         Ling Fu        \and 
%         Qian Liu       \and
%         Gang Zheng 
%          }

\authorrunning{Yun  et al} % if too long for running head

\institute{      
           Ling Fu(Corresponding Author), Qian Liu, Qiuhao Wang  \at
              School of BioMedical Engineering, Hainan University, Haikou, 570288, China,
           \and
           Gang Zheng, Enze Zhou  \at
              School of Electronic Information and Communications, Huazhong University of Science and Technology, Wuhan, 430074, China,
            \and
            Yun Chu \at
              School of Information and Communication, Hainan University, Haikou, 570288, China  
}

\date{Received: date 2024.04 / Accepted: 2025.01}

\maketitle
\begin{abstract}
Auscultation plays a pivotal role in early respiratory and pulmonary disease diagnosis. Despite the emergence of deep learning-based methods for automatic respiratory sound classification post-Covid-19, limited datasets impede performance enhancement. Distinguishing between normal and abnormal respiratory sounds poses challenges due to the coexistence of normal respiratory components and noise components in both types. Moreover, different abnormal respiratory sounds exhibit similar anomalous features, hindering their differentiation. Besides, existing state-of-the-art models suffer from excessive parameter size, impeding deployment on resource-constrained mobile platforms. To address these issues, we design  a lightweight network CycleGuardian and  propose a framework based on an improved deep clustering and contrastive learning.  We first generate a hybrid spectrogram for feature diversity and grouping spectrograms to facilitating intermittent abnormal sound capture.Then, CycleGuardian integrates a deep clustering module with a similarity-constrained clustering component to improve the ability to capture abnormal features  and a contrastive learning module with group mixing for enhanced abnormal feature discernment. Multi-objective optimization enhances overall performance during training. In experiments we use the ICBHI2017 dataset, following the official split method and without any pre-trained weights, our method achieves  Sp: 82.06 $\%$, Se: 44.47$\%$, and Score: 63.26$\%$ with a network model size of 38M, comparing to the current model, our method leads by nearly 7$\%$, achieving the current best performances.
Additionally, we deploy the network on Android devices, showcasing a comprehensive intelligent respiratory sound auscultation system.

\end{abstract}

\section{Introduction}
According to the World Health Organization (WHO), the five major respiratory diseases \cite{world2022world}–lung cancer, tuberculosis, asthma, acute lower respiratory tract infections (LRTI), and chronic obstructive pulmonary disease (COPD) collectively result in over 3 million deaths worldwide annually \cite{mathers2006projections} \cite{harding2020global}. Pulmonary diseases have become the third leading cause of death globally\cite{forum2017global}. These respiratory ailments significantly impact healthcare systems and adversely affect people's lives. Prevention, early diagnosis, and treatment are recognized as key factors in reducing the adverse effects of these deadly diseases.

Pulmonary auscultation\cite{bohadana2014fundamentals} is a crucial method for examining respiratory system diseases. By listening to lung sounds, experts can identify adventitious sounds during the respiratory cycle, including crackles and wheezes. These non-periodic and non-stationary sounds are primarily categorized into two groups: normal (vesicular) and abnormal (adventitious)\cite{pramono2017automatic}. Normal respiratory sounds are observed in the absence of respiratory diseases, with frequencies ranging from 100 to 2000 Hz. Abnormal respiratory sounds can be further classified as continuous or discontinuous, with wheezes and crackles being the most common types\cite{reichert2008analysis}, respectively. Wheezing is a significant symptom of asthma and COPD, with its main frequency concentrated around 400 Hz and a duration exceeding 80 ms. In contrast, crackles are discontinuous anomalies often associated with obstructive pulmonary diseases, including COPD, chronic bronchitis, pneumonia, and pulmonary fibrosis. Crackles comprise fine and coarse crackles, with fine crackles having a frequency of approximately 650 Hz and a duration of around 5 ms, while coarse crackles have a frequency of approximately 350 Hz and a duration of about 15 ms. These abnormal respiratory sounds can be characterized based on frequency, pitch, energy, and differentiation from normal lung sounds.

However, even for experts, identifying differences between various abnormal respiratory sounds poses challenges, as different experts may introduce subjectivity in diagnostic interpretations\cite{bahoura2003new}. Furthermore, in many underdeveloped regions lacking experienced physicians, delays in disease diagnosis and treatment occur. In such circumstances, people have begun using stethoscopes to collect respiratory sounds, followed by computer-aided automatic classification of these sounds. Existing research primarily focuses on addressing the following aspects: dealing with class imbalance in respiratory sound data, feature extraction, and network  design.

\begin{figure*}[htbp]
\centering
\includegraphics[width= 1.0\textwidth]{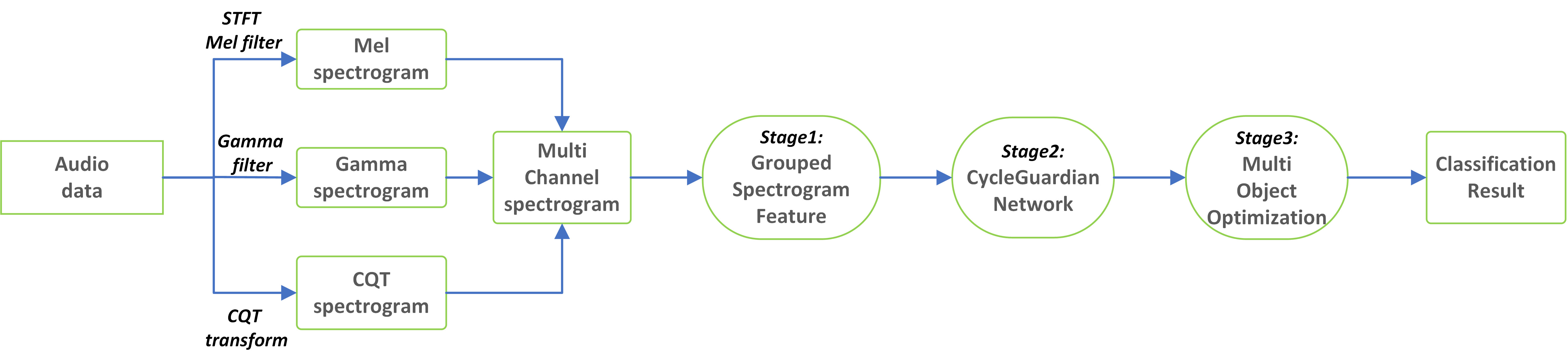}
\caption{Flowchart of the automatic respiratory sounds classification framework. In stage 1, the obtained multichannel speech maps are grouped. In stage 2, feature encoding of each group of speech sounds is done by CycleGuardian network, and deep clustering and comparison learning is performed. In stage 3, joint optimisation of multiple objectives is performed.}
\label{Fig.1}
\end{figure*}

Class imbalance, prevalent in medical datasets, presents a scarcity of abnormal samples compared to normal ones, a pattern also evident in respiratory sound datasets. \cite{gairola2021respirenet} addresses this by employing data augmentation, concatenating audio samples to generate new abnormal instances. Similarly, \cite{ma2020lungrn+} utilizes mixup techniques on spectrogram features to augment abnormal data at the feature level.   \cite{petmezas2022automated} introduces focal loss to mitigate class imbalance by assigning higher weights to hard samples.\cite{kim2023adversarial} utilized generative models to synthesize abnormal data and applied adversarial fine-tuning techniques to align the features between the generated and real data, addressing the issue of data imbalance. \cite{kim2024repaugment} proposed an audio-level enhancement technique to improve the feature representation.  \cite{chua2024towards}  introduced a multi-label learning method to handle the scarcity of abnormal samples.

 In feature extraction from respiratory sounds, some methods decompose the signal into  subsequent signals. \cite{li2021lungattn} decomposes audio signals into three components, extracting spectrograms from each with different window sizes. \cite{shuvo2020lightweight} employs empirical mode decomposition to derive intrinsic mode function coefficients, then applies wavelet transform for feature extraction. \cite{gupta2022classification} combines respiratory sounds with lung volume using MVMD to predict diseases.On the other hand, feature extraction directly from the original respiratory sound signal is a more common approach in research. \cite{wang2024ofgst}, \cite{chen2019triple} improves the S-transform to obtain spectrograms, while \cite{minami2019automatic} and \cite{ma2019lungbrn} utilize STFT and WT simultaneously for spectrogram extraction.\cite{xu2021arsc} and \cite{bardou2018lung} directly extract MFCC coefficients and MEL spectrograms. \cite{gupta2021gammatonegram},\cite{mang2024classification} employs gammatone filter banks, and \cite{li2010feature} analyzes second-order spectral function to obtain bispectrogram features. Direct spectrogram extraction simplifies feature extraction, enabling end-to-end classification with neural networks. However, both normal and abnormal samples contain common information, predominantly normal components and noise. Extracting spectrograms in image form hinders the network's ability to discern abnormal information effectively. Early classification relied on empirical thresholds, evolving to  deep neural networks\cite{mukherjee2021automatic}, \cite{demir2019convolutional},\cite{liumasked}, often utilizing pre-trained weights for transfer learning. Challenges arise in distinguishing complex symptom abnormal features from single symptom ones due to their strong similarities.

In summary, for the task of abnormal respiratory sound classification, the key issues that need to be addressed are as follows: 1.There exists class imbalance among different categories in the dataset. 2. Distinguishing between normal and abnormal respiratory sounds presents a challenge due to the concurrent presence of normal respiratory components and noise components in both categories. 3. Different types of abnormal respiratory sounds often share similar anomalous features, particularly between complex symptomatic patterns and single-symptom presentations, further complicating their discrimination. What's more, the utilization of current state-of-the-art models\cite{bae2023patch} is hindered by excessive parameter and model size, impeding their deployment\cite{cai2023efficientvit}, \cite{huang2023deep} on resource-constrained mobile platforms.

To address these issues, we propose a method based on deep clustering and contrastive learning and design a lightweight network  specifically for handling grouped respiratory spectrogram, named CycleGuardian. {Fig.1} illustrates the overall design concept of this paper. The framework mainly consists of three components: 1) Grouped spectrogram feature generation. 2) Deep clustering and contrastive learning. 3) Multi-loss optimization strategy. The workflow of the entire framework is described below.

We first group and encode the spectrogram features to capture intermittent abnormal respiratory sounds effectively. Unlike patching methods\cite{song2023patch}, our method directly groups spectrograms in the time dimension at the frame level, as illustrated in Stage 1 of {Fig.1} (Section 3.1). Following the encoding of grouped features,  we propose an improved deep clustering approach to cluster these features and introduce a constraint loss based on cosine similarity to reduce similarity between cluster features. On the other hand, to enhance discriminability between complex and single abnormal features, we introduce a contrastive learning branch, this branch randomly replaces a portion of grouped features with features from other samples in the batch, facilitating contrastive learning between original and mixed samples. We employ a contrastive loss based on the similarity between global vectors ($glo_{ori}$ and $glo_{mix}$) (Section 3.2). Furthermore,  to optimize network performance, we utilize joint optimization with multiple objective losses. Different optimizers and weights are assigned to these losses for joint optimization (Section 3.3).

To the best of our knowledge, the work to combine deep clustering with contrastive learning  and apply it to respiratory  sound  classification is currently rarely explored.  The main contributions of this paper are summarized as follows: 

\begin{itemize}

\item [1)]
Comparing  to conventional methods of directly encoding spectrograms or applying patch-based techniques, we explore the approach of grouping spectrogram features, which is advantageous for capturing abnormal  information from normal respiratory sounds, and augment the network's ability to differentiate between normal and abnormal respiratory sounds.

\item [2)]
We design a lightweight network architecture model, CycleGuardian, and propose an improved method of deep clustering and contrastive learning. By combining them, the network's capacity to distinguish between different abnormal respiratory sounds is improved. During training, we introduce a multi-objective loss joint optimization strategy to simultaneously optimize the clustering module, contrastive learning, and classification module, thus enhancing the overall performance of the network.

\item [3)]
Extensive experiments on the ICBHI2017 dataset demonstrate the effectiveness and superiority of our method. CycleGuardian achieves state-of-the-art performance without pretrained weights, utilizing a compact architecture of only 38MB model size. Additionally, we conduct preliminary deployment on smartphones, paving the way for an intelligent respiratory sound diagnosis system.

\end{itemize}

\section{Related work}
\subsection{Contrastive learning}

Contrastive learning is a self-supervised learning technique that learns effective feature representations by contrasting positive and negative pairs without the need for explicit data labeling, which is particularly important when labeled data is scarce or costly to obtain. It enables models to extract feature representations from the structure of the data itself, which can be applied to various downstream tasks such as classification, detection, and segmentation.

Recent work has applied contrastive learning to abnormal classification tasks in respiratory sounds. Combining contrastive learning with multiple instance learning, studies \cite{song2023patch} \cite{song2021contrastive} divide each sample into multiple patches, categorizing patches into different subclasses based on the types of respiratory sounds they contain. Since patch-level labels are often unavailable, they estimate patch-level labels using a multiple instance learning (MIL) approach, minimizing the distance between features of patches from the same subclass and maximizing the distance between features of patches from different subclasses, forming a contrastive loss for learning. Another paper \cite{bae2023patch}  proposes the use of pre-trained weights on large-scale visual and audio datasets and introduces a Patch-Mix contrastive learning method to improve respiratory sound classification performance. Combining metadata and labels, study \cite{moummad2023pretraining} proposes a supervised contrastive learning method to extract useful representations of respiratory sounds, demonstrating that using supervised contrastive learning combined with metadata learning is superior to cross-entropy training. \cite{kim2024stethoscope} suggested a stethoscope-guided contrastive learning method to reduce the model's dependency on stethoscope data

\subsection{Deep Clustering}
Clustering aims to group given objects such that the similarity among objects within the same cluster is higher than those in other clusters. It is an unsupervised learning method widely applied in various practical applications like image classification and data visualization. Traditional clustering methods such as K-means (KM), Gaussian Mixture Model (GMM), and Spectral Clustering (SC) are known for their speed and broad applicability. However, these methods often rely on predefined similarity measures, which become ineffective when dealing with high-dimensional data. To address this issue, researchers have employed dimensionality reduction techniques to project the original data into a lower-dimensional space before clustering, thereby facilitating effective clustering. Nevertheless, the limited expressive power of low-dimensional features restricts the representation ability of the data.

Compared to traditional methods, deep neural networks (DNNs) excel in extracting high-level features through nonlinear embeddings, benefiting clustering tasks. Approaches like Deep Embedded Clustering (DEC) \cite{xie2016unsupervised} and Deep K-means (DKM) \cite{fard2020deep} typically employ a two-stage learning process, separating feature extraction and clustering. However, this separation overlooks potential relationships, limiting performance. To address this, researchers explore joint unsupervised learning \cite{yang2016joint} and propose models like deep adaptive clustering (DAC) \cite{chang2017deep} and Adaptive Self-Paced Clustering (ASPC) \cite{guo2019adaptive}. These models prioritize high-confidence samples and incorporate constraints into deep clustering processes to enhance performance. Constraints include pairwise, instance hardness, triplet, and cardinality constraints \cite{zhang2021framework}. Additionally, instance-level must-link and cannot-link constraints \cite{gonzalez2020agglomerative} leverage pairwise distances between instances to enhance constrained clustering capability.

Inspired by the aforementioned works, this paper integrates deep clustering with contrastive learning and bring them into the task of respiratory sound classification. Unlike previous studies, we introduce a cosine similarity-based constraint loss in deep clustering to further reduce inter-cluster correlations and enhance clustering performance. In contrastive learning, we modify the encoding of spectrograms by grouping them directly at the frame level instead of patch encoding. The features of different samples within the same batch are mixed using group mix to generate new mixed samples for contrastive learning.

\section{Proposed method}
In this section, we introduce the generation of multi-channel spectrogram features and their grouping encoding to obtain group features. Subsequently, we present the two main components of CycleGuardian: the deep clustering module for clustering group features into different clusters, and the contrastive learning module, where group features from different samples are mixed to generate new mixed sample features for contrastive learning. Finally, we perform joint optimization of deep clustering and contrastive learning, detailing the proposed method.

\subsection{Multi Channel Spectrogram Grouping}
To enhance the diversity of spectrogram features, we construct multi-channel spectrogram features consisting of Mel spectrograms, CQT, and gamma spectrograms.

\subsubsection{Multi Channel Spectrogram Generation }

In this method, Mel spectrograms and gamma spectrograms are generated by designing respective filters based on the cochlear effect, providing higher resolution for low-frequency components and lower resolution for high-frequency components. They are both formed based on Fast Fourier Transform (FFT). Given a fixed-length audio segment $x[n]$, the FFT is used to transform the time-domain signal into the frequency domain. Mel spectrograms extract frequency bands through a set of Mel filters and scale the power spectrum logarithmically to form Mel spectrograms, denoted as $S_0$. The conversion between Mel frequency and Hertz frequency is given by $f_{\mathrm{Mel}}=2595\log_{10}\left(1+{\frac{f}{700}}\right)$. Gamma spectrograms, denoted as $S_1$, are formed by applying different enhancements to different frequency bands of the frequency-domain signal using gamma-tone filter banks and logarithmically scaling the filter outputs.

Unlike Mel and gamma spectrograms, Constant Q Transform (CQT) is another method of transforming signals into time-frequency domains, providing a constant Q factor as frequency varies. Here, $Q={\frac{f_{\mathrm{k}}}{\Delta f_{\mathrm{k}}}}$, where $f_{\mathrm{k}}$ is the center frequency of the $k$-th bin and $\Delta f_{\mathrm{k}}$ is the bandwidth. Each bin's frequency signal is obtained through $X[k]_{cqt}=\sum_{n=0}^{N-1}x[n]\cdot e^{-{\frac{2\pi i}{Q}}{\frac{k}{f_{s}}}n}$, followed by a logarithmic transformation to obtain the CQT spectrogram, denoted as $S_2$.

After obtaining the three different types of spectrograms, we generate multi-channel spectrogram features using the following equation \eqref{1}.

\begin{equation}
S = S_0 \oplus   S_1  \oplus   S_2 , \label{1}\\
\end{equation}

Here, $S \in R^{c \times m \times n}$ represents the three channels of spectrograms generated, all of which are $m \times n$ matrices. Here, $m$ represents the frequency dimension, $n$ represents the number of frames, and $c$ represents the number of channels. By using the $\oplus$ operation, we concatenate them along the channel dimension to generate a three-channel spectrogram feature.

\subsubsection{Network Basic Unit }

As shown in {Fig.2}, four basic units are employed in our network model. The CBR unit consists of Convolution, BatchNorm, and ReLU activation functions. The CGL unit comprises convolution, GELU, and LayerNorm. LGL is composed of Linear, GELU, and LayerNorm. The Group Feature Embedding (GFE Unit) module cascades  with three CBR unit and the linear layers. Using these basic units, we construct the improved Deep Embedding Cluster module (IDEC) and the Cluster Projection Fusion Module (CPF Module), forming the entire architecture of CycleGuardian.

\begin{figure*}[htbp]
\begin{center}
\includegraphics[width=0.7\textwidth]{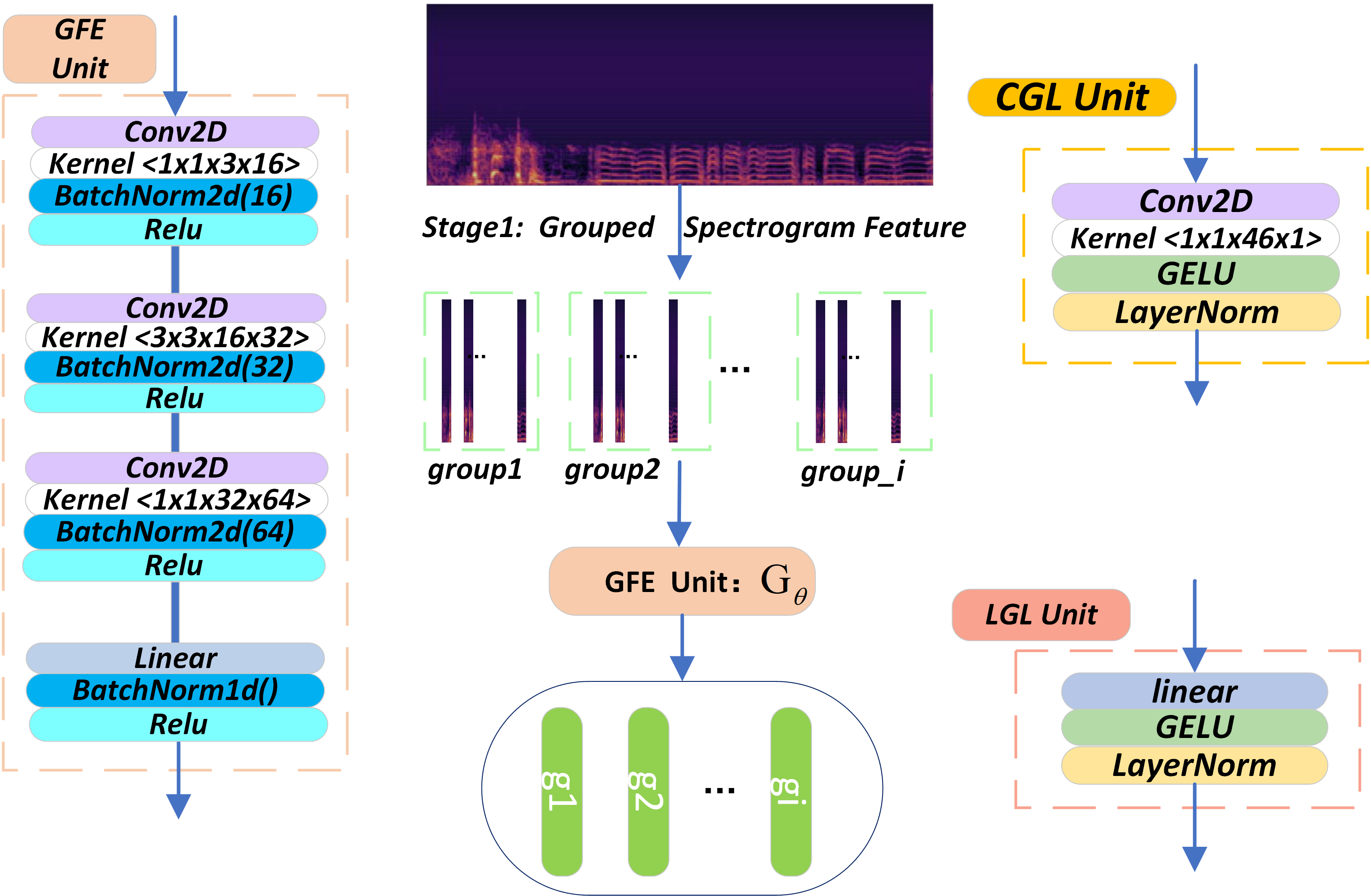}    
\end{center}
\caption{The left part shows the  GFE Unit in the network,  the middle shows the idea of grouping and encoding the spectrogram, and the right part shows the CGL, LGL unit. }
\label{Fig.2}
\end{figure*}

\subsubsection{Group Feature Embedding}

The right side of {Fig.2} illustrates the concept of group encoding. Initially, the multi-channel hybrided spectrogram features $S \in R^{c \times m \times n}$ are grouped, with each group containing information from 20 frames of spectrograms. These grouped spectrograms are then inputted into the Group Feature Embedding (GFE Unit) module to obtain encoded features $g_{i}$ for each group. Group encoding of spectrograms enhances the capturing ability for intermittent abnormal respiratory  sound such as crackles.

\begin{equation}
\begin{aligned}
& g_{i} = G_{\theta}(group_{i}),  i \in {0,1,...(N_{g}-1) },\\
& G_{\theta}(\bullet)= GFE_{unit}(\bullet), \label{2}
\end{aligned}
\end{equation}

In \eqref{2}, $group_{i}$ represents the grouped spectrograms, $G_{\theta}(\bullet)$ denotes the GFE unit, ${\theta}$ represents the learnable parameters in the module, and $g_{i}$ represents the output of the group encoding module used to represent group encoding features. Let $T_{frames}$ denote the total number of frames in each sample. By sliding $S_{frames}$ frames to the right each time, the number of groups obtained for each sample is denoted as $N_{g} = \frac{T_{frames}}{S_{frames}}$. Additionally, the sliding frames $S_{frames}$ must be less than the number of frames included in each group, $G_{frames}$, to ensure approximately 25 $\%$ overlap between each grouped spectrogram.

\subsection{Improved Contrastive Learning and Deep Clustering}
In this section, we first introduce the contrastive learning module based on group mix, which blends group features from each sample to generate new mixed sample features for contrastive learning against the original samples. Next, we elucidate the deep clustering module based on similarity constraints, used to cluster multiple group features from each sample into five clusters. Subsequently, we provide an overview of the overall architecture and operation flow of CycleGuardian.

\subsubsection{Contrastive learning based on group mix}

Considering that directly mixing the original audio may result in mislabeling of audio samples containing short segments of crackles or wheezes, the labels of the newly generated audio samples should ideally be marked as crackle or wheeze. However, since we cannot directly determine whether the mixed-in audio segment is normal or abnormal during the mixing process, making it difficult to assign labels to the newly generated samples, therefore, our study elects to implement group mixing at the feature level.

We proposed a contrastive learning based on group mix,  as shown in {Fig.3}, each sample's spectrogram is divided into multiple groups, with each group containing 20 frames of spectrogram information. These grouped spectrograms are then inputted into the Group Feature Embedding (GFE)  Unit  to obtain encoding features $g_{i}$ for each group. Within a batch, each sample generates multiple group features $g_{i}$. By using group mix, new mixed sample features are generated. The proportion of original sample's group features retained in the mixed sample feature is denoted as $\lambda$, with the replacement ratio being $1-\lambda$. Based on this approach, contrastive learning is performed between the original samples and the mixed samples.

\begin{figure*}[htbp]
\begin{center}
\includegraphics[width=0.41\textwidth]{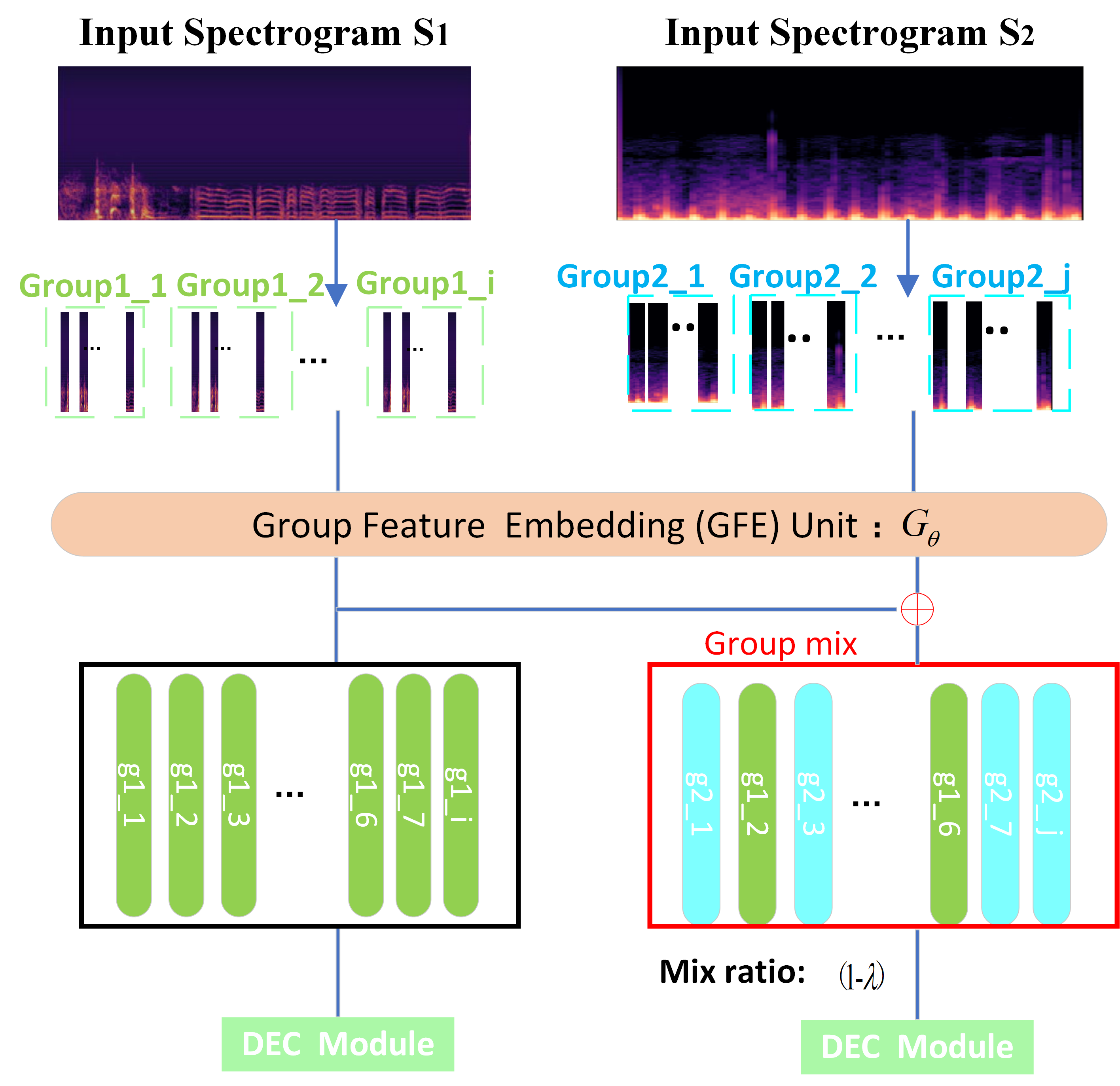}    
\end{center}
\captionsetup{font=large} % Adjust font size here
\caption{Group mix module: input group$_i$ into GFE Unit to get group feature g$_i$, and then use g$_i$ in group mix to generate new hybrid sample features for subsequent contrastive learning.}
\label{Fig.3}
\end{figure*}

As shown in equation  \eqref{3},  $z_{i}$ represents the global feature of the current original sample, while $\hat z_{i}$  denotes the global feature of the current sample after group mix. Since in one batch contains multiple samples,  $z_{m}$ refers to the global feature generated by other samples and is also another part groups of the source for generating $\hat z_{i}$  . The similarity between $\hat z_{i}$ and $z_{i}$, as well as between $\hat z_{i}$ and  $z_{m}$, is controlled by the parameter $\lambda$ and $1-\lambda$. The main idea behind this contrastive learning is that if the mixed sample feature retains the proportion of original sample group features as $\lambda$, then theoretically, the similarity between the mixed global feature and the original global features should also be approximately $\lambda$.

\begin{equation}
L_{\mathrm{con}}=-\ \frac{1}{\vert I\vert}\sum_{i\in I}\left[\left(\lambda\cdot\left(h(\hat{z}_{i})^{\top}z_{i}/\tau\right)+(1-\lambda)\cdot\left(h(\hat{z}_{i})^{\top}z_{m}/\tau\right)\right) \right], \label{3}\\
\end{equation}

Here, $h(\bullet)$ is a mapping module composed of two MLP layers with ReLU and BN layers. $I$ represents the number of samples contained in one batch, $\tau$ controls the sharpness of cosine similarity, and $\lambda$ is sampled from the $\beta$ distribution. Additionally, all global features, such as $z_{i}$ or $h(\hat z_{i})$, are normalized before the dot product.

\subsubsection{Improved Deep clustering based on similarity constraints}

In traditional clustering algorithms like K-means, each data point or feature is hard-assigned to a single cluster. However, in many real-world scenarios, data points or features may belong to multiple clusters or reside in fuzzy regions between clusters, such as complex symptoms containing both abnormal features like crackles and wheezes. To address this limitation, we introduce the idea of Deep Embedding Clustering (DEC) in our approach, which employs a soft assignment strategy. This soft assignment is implemented using a Student's t-distribution, where each group encoding feature is associated with a probability distribution over clusters.

In equation \eqref{4}, $\mathbf{e}_i$ represents the group encoding feature generated after dimensionality reduction by an autoencoder. $\boldsymbol{\mu}_j$ is the centroid of cluster j. $q_{ij}$ models the relationship between $e_{i}$  and  $\mu_{j}$. Considering the high dimensionality and nonlinearity of the embedding space, directly optimizing $q_{ij}$ may be challenging. Hence, an auxiliary target distribution $p_{ij}$ is introduced, as shown in equation \eqref{5}. During training, $p_{ij}$ serves as a reference for $q_{ij}$, providing a more stable and effective optimization process for the entire clustering.

\begin{equation}
q_{ij} = \frac{{(1 + ||\mathbf{e}_i - \boldsymbol{\mu}_j||^2 / \alpha)^{-\frac{\alpha + 1}{2}}}}{{\sum_{j'=1}^{k}(1 + ||\mathbf{e}_i - \boldsymbol{\mu}_{j'}||^2 / \alpha)^{-\frac{\alpha + 1}{2}}}}, i \in I \label{4}\\
\end{equation}

\begin{equation}
p_{ij}=\frac{q_{i j}^{2}}{\sum\limits_{i}q_{i j}}\left(\sum_{j^{\prime}}\,\left(\frac{q_{i j^{\prime}}^{2}}{\sum\limits_{i}q_{i j^{\prime}}}\,\right)\right)^{-1}, \label{5}\\
\end{equation}

Here, $i \in I$ represents the number of group features contained in each sample  and it's value range from  $[0, N_{g}-1]$, k is the total number of clusters. $\alpha$ denotes the degrees of freedom of the Student's t-distribution, controlling the concentration of the distribution.  In DEC, $\alpha$ is typically set to a predefined value (e.g., 1), but it can also be learned. $q_{ij}$ is termed as soft assignment, interpreting the probability of the clustering module assigning the i-th group encoding feature $\mathbf{e}_i$ to the j-th cluster.

In contrast to previous Deep Embedding Clustering(DEC) approaches, we propose an Improved Deep Embedding Clustering (IDEC) to enhance the differentiation between cluster features. The principle is to introduce a constraint based on cosine similarity on top of DEC during the clustering process. As illustrated in {Fig.4},the IDEC include two part which are DEC module and Cluster Projection Fusion (CPF) module. In the DEC module, the group encoding features $g_{i}$ are  dimensionally reduced to $\mathbf{e}_i$, the number of clusters is set to five which are representing normal, noise, and three types of abnormal features, through clustering, each $e_{i}$ is assigned to the cluster with the highest probability.  After determining each group's cluster assignment, the CPF module maps all group features belonging to the same cluster into a cluster feature $c_{i}$. It's worth to note that the cluster features $c_{i}$ are generated using the group encoding features $g_{i}$ before dimensionality reduction. After obtaining the cluster features for each cluster, these five cluster features are fused into a global feature  $z$ through weighted fusion.

\begin{figure*}[htbp]
\begin{center}
\includegraphics[width=0.7\textwidth]{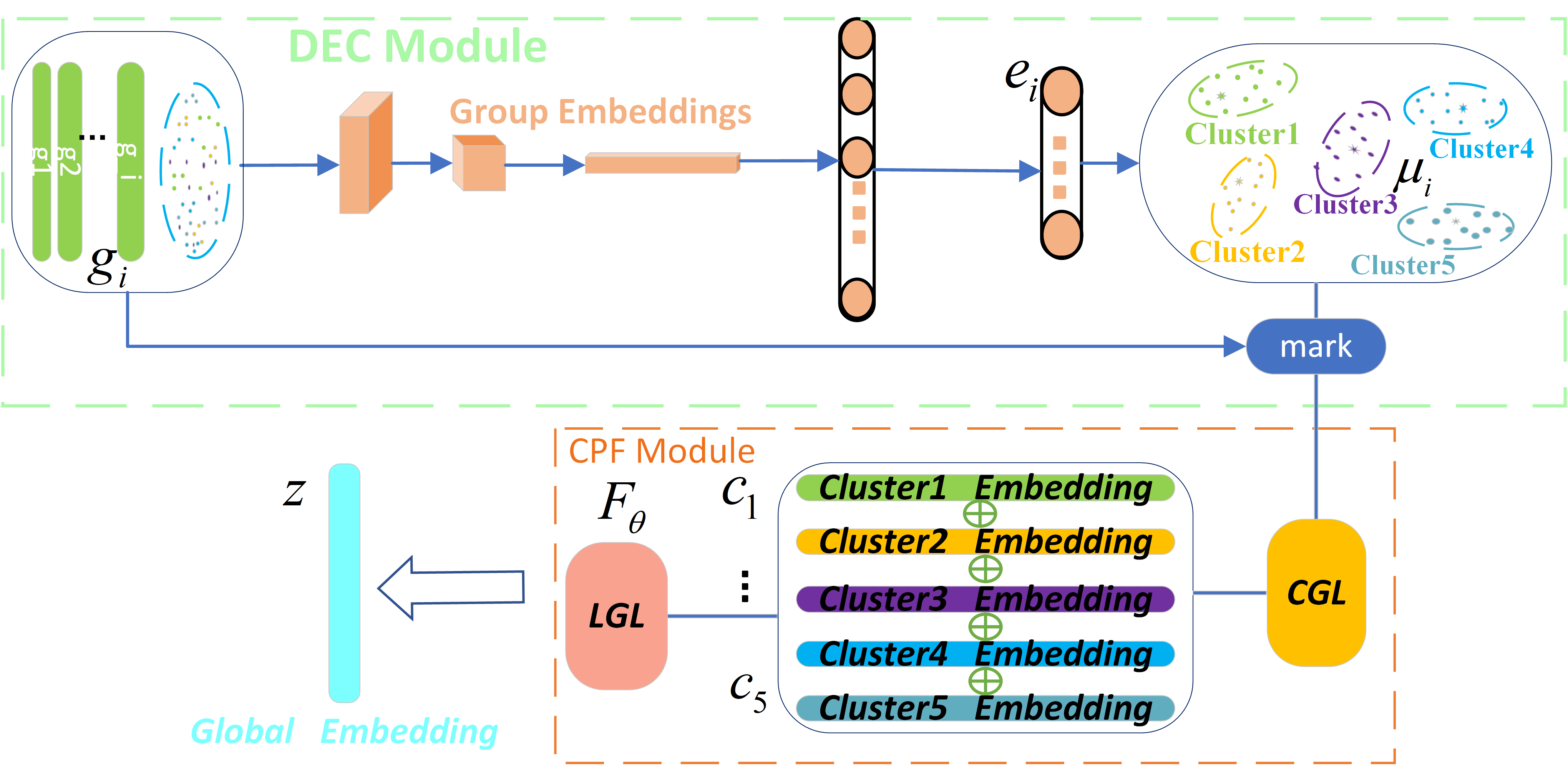}    
\end{center}
\caption{The  Improved Deep  Embedding Clustering (IDEC) Module: including the DEC Module and Cluster Projection Fusion (CPF) Module.}
\label{Fig.4}
\end{figure*}

As defined in equation \eqref{6} , the similarity between any two cluster features $c_{i}$ is quantified using soft cosine similarity, reducing the correlation between these 5 cluster representation features, forming a constrained deep clustering process.

\begin{equation}
\begin{aligned}
& AbsSoftCos_{sim}(c_{i}, c_{j}) =  \left| \frac{c^T_i S c_j}{ \sqrt{(c^T_i S c_i)(c^T_j S c_j)} } \right|, \\
& c_{i} = F_{\theta}(g_{i}), i,j \in[1,..k], \label{6}\\
\end{aligned}
\end{equation}

Here, in the numerator,  $c_{i}^T S c_{j}$ representing the weighted dot product between the features $c_{i}$ and $c_{j}$, where the weights are provided by the similarity matrix $S$. In the denominator, $\sqrt{(c_{i}^T S c_{i})(c_{j}^T S c_{j})}$ is the product of the norms of the feature $c_{i}$ and $c_{j}$, adjusted by the similarity matrix  which is   composed by learnable parameters. $F_{\theta}(\bullet)$ represents the cluster feature fusion (CPF) module, where $\theta$ denotes the learnable parameters of this module.  The use of absolute value $\left|\right|$ is due to the cosine similarity values ranging from (-1,1). Since this similarity will be used as a loss term, taking the absolute value helps prevent positive and negative values from canceling each other during optimization.

\subsubsection{Network Architecture}
The network architecture of CycleGuardian, as depicted in {Fig.5}, operates as follows. Firstly, in Stage 1, the spectrograms of each sample are divided into multiple groups. Secondly, in Stage 2, each group's spectrogram is input into the GFE unit to obtain each group's encoding feature $g_{i}$. Next, all the $g_{i}$ from the same batch are fed into two branches. In the lower branch, $g_{i}$ undergoes the group mix module, which replaces the $g_{i}$ from the original sample with those from other samples at a certain ratio, generating new mixed sample features input into the DEC module. The DEC module first reduces each $g_{i}$ to $e_{i}$, then clusters all $e_{i}$ into five clusters, ensuring similar $e_{i}$ are assigned to the same cluster. Subsequently, in the CPF module, all group encoding features $g_{i}$ belonging to the same cluster are mapped into one cluster feature $c_{i}$. After generating five cluster features  $c_{i}$, they are fused into a tensor $\hat z_{i}$ using adaptive weighting, representing the global feature formed by the mixed sample features.

\begin{figure*}[htbp]
\centering
\includegraphics[width= 0.8\textwidth]{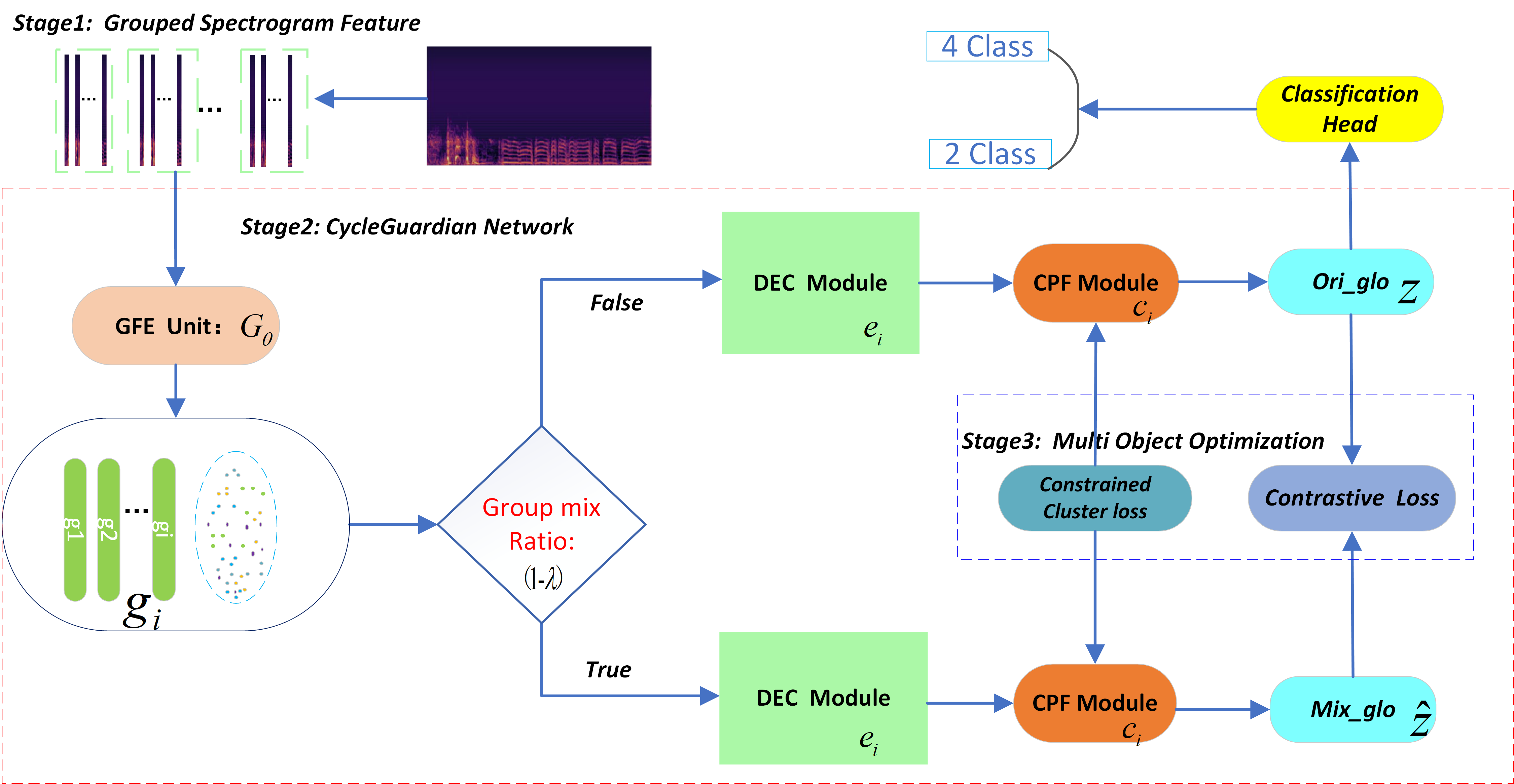}
\captionsetup{justification=raggedright,singlelinecheck=false} % Set caption justification to left-aligned
\caption{%
%\begin{minipage}{2\columnwidth}
CycleGuardian network architecture: the red dashed box contains the main modules of the network, the GFE Unit, Group mix, DEC, and CPF module. The blue dotted box contains the loss terms to be optimised, the clustering loss based on similarity constraints, the comparison loss, and the classification loss.
%\end{minipage}
}

\label{Fig.5}
\end{figure*}

Meanwhile, in the  upper branch, the group encoding features $g_{i}$ bypass the group mix operation and are directly input into the DEC module. They then pass through the CPF module to generate the original sample's global feature $z_{i}$. The DEC and CPF modules in both branches share the same weight parameters. In stage 3, the newly generated mixed sample feature obtains a global feature $\hat z_{i}$ and the original sample generates a global feature $z_{i}$. Contrastive loss is constructed between the global features of mixed samples $\hat z_{i}$ and the global features of other samples in the same batch.

\subsection{Multi object loss optimization}

At this point, three types of objectives need to be optimized in the entire framework: the contrastive learning module based on group mix, the deep embedding clustering module, and the final classification module. In the previous section, we have elaborated on the composition of the loss in the contrastive learning module. Here, we will discuss the composition of the clustering loss and classification loss.

The clustering loss mainly consists of two aspects. Firstly, using the soft assignment $q_{ij}$ and the auxiliary target distribution $p_{ij}$, we construct a clustering objective loss based on the KL divergence, denoted as $L_{clu}$, as shown in  \eqref{7}. On the other hand, we impose a constraint based on soft cosine similarity to reduce the similarity between any pair of the five cluster features, $cos_{sim}(c_{i}, c_{j})$ representing the cluster vector similarity loss $L_{cos}$, as shown in  \eqref{8}:

\begin{equation}
L_{clu}=\mathrm{\sum\limits_{i=1}^{I}}\sum\limits_{j=1}^{k}  p_{i j}\mathrm{ln}\left(\frac{p_{i j}}{q_{i j}}\right), \label{7}\\
\end{equation}

\begin{equation}
L_{cos}= \sum\limits_{i=1}^{k} \sum\limits_{j=1}^{k}  AbsSoftCos_{sim}(c_{i}, c_{j}), \label{8}\\
\end{equation}

Where $(c_{i}, c_{j})$ represents the generated cluster vector representations, $k$ denotes the total number of clusters, set to 5 in this paper. Additionally, for the classification loss between predictions and labels, we employ cross-entropy as the loss function, denoted as $L_{cls}$, as shown in  \eqref{9}:

\begin{equation}
L_{cls} =-\frac{1}{N}\sum\limits_{n=1}^{N}\sum\limits_{c=1}^{C}y_c^{(n)}log\hat y_c^{(n)}, \label{9}\\
\end{equation}

Where $N$ represents the number of samples, $C$ represents the number of classes, $y_{c}$ and $\hat y_{c}$ represent the original class labels and predicted labels, respectively. With this, we have constructed the components of the entire framework's optimization objectives, which include the contrastive loss, clustering loss based on soft  cosine similarity constraints, and classification loss. The overall loss is denoted as $L_{total}$, as shown in  \eqref{10}:

\begin{equation}
L_{total} = L_{con}+ \alpha L_{clu} + \gamma L_{cos}+ L_{cls}, \label{10}\\
\end{equation}

Where $\alpha$ and $\gamma$ are hyperparameters used to balance the weights of clustering and similarity losses in the overall loss function. $L_{con}$, $L_{clu}$, $L_{cos}$, and $L_{cls}$ represent the contrastive loss, clustering loss, cluster similarity constraint loss, and classification loss, respectively, as discussed earlier.

\section{Experiments}

In this section, we provide a detailed description of the dataset, experimental setup, training process, and parameter configurations. We present qualitative and quantitative experimental results along with analysis.

\subsection{Dateset  and Environment}
Our study is based on the ICBHI2017 dataset, the largest open-source dataset available to date for respiratory sound analysis. The majority of research on respiratory sounds is conducted using this dataset, facilitating experimental comparisons. The dataset comprises 920 audio files and corresponding annotated texts from 126 participants, including adults and children. Each audio recording is segmented based on annotated time intervals to extract respiratory sound cycles. Consequently, a total of 6898 respiratory sound cycles are generated. The distribution of each category is outlined in {Table1}.

According to {Table1}, the dataset exhibits class imbalance issues. To address this, we perform data augmentation at two levels: raw audio and spectrogram features. We alleviate this problem by applying different data augmentation techniques randomly to each batch during training.

\begin{table}[htbp]
\flushleft
\captionsetup{justification=raggedright}
\caption{Number of samples and ratios  for four class on trainset and validset}

\setlength{\tabcolsep}{2mm}{
\begin{tabular}{lccccccccc} % 控制表格的格式
\toprule
% \multirow{1}{*} Model Number   \\
% \cline{2-6}  % 这部分是画一条横线在2-6 排之间
 Class &  normal   & crackle &    Wheeze   &  both & Total  \\
\midrule
  Number     &     3642   &  1864 &  886  & 506    & 6898   \\
  $Train_{set}$  &     0.49   &  0.29 &  0.12 & 0.10   &  4262   \\
  $Valid_{set}$  &     0.55   &  0.24 &  0.14 & 0.07   &  2873  \\
\bottomrule
\end{tabular} 
    }
\flushleft
  \label{Table1}
\end{table}

For ease of comparison with other works, we utilize the evaluation metrics provided by the official ICBHI evaluation function. Sensitivity $(S_e)$ and Specificity $(S_p)$ are employed, with the final score being the average of sensitivity and specificity, calculated as shown in \eqref{11}:

\begin{equation}
S_e = \frac{P_c +P_w + P_b}{N_c+ N_w + N_b},S_p = \frac{P_n}{N_n},S_{score} = \frac{S_e + S_p}{2}, \label{11}\\
\end{equation}

Here, $P_i$ denotes the number of correctly classified samples in category $i$, and $N_i$ represents the total number of samples in category $i$, the classes include four categories: $i\in normal,crackle, wheeze,both$.

Additionally, the experiments are conducted on Ubuntu 20.04, utilizing Python 3.8 within an Anaconda3-created virtual environment. We employ PyTorch 2.0.1 as the deep learning framework with CUDA 11.2. The hardware setup consists of an Intel Core i9 processor, 24GB RAM, and an Nvidia GeForce RTX 3090 with 24GB memory.

\subsection{Data pre processing and parameter setting }

\subsubsection{Data pre processing }

\textbf{Sample Length Alignment}: The original dataset comprises over 6800 samples with varying durations ranging from 0.2s to 16.2s. Approximately 65 $\%$ of respiratory sound cycles are less than 3s, 33$\%$ have lengths between 4-6s, with an average duration of 2.7s. To ensure uniformity in the length of respiratory sounds used in experiments and considering a typical respiratory cycle lasts 2-3s, we follow the approach suggested by \cite{gairola2021respirenet} to fix all respiratory sound cycles to 8s. For samples shorter than 8s, we randomly extract segments from the sample itself and concatenate them to reach a length of 8s. If a sample exceeds 8s, we truncate it to retain only the initial 8s.

\textbf{Data Augmentation}: As mentioned earlier, to address data imbalance, data augmentation is performed at two levels: raw audio and spectrogram features. During preprocessing, various augmentation techniques such as adding noise, shifting, stretching, and VTLP are randomly applied to audio samples within each batch. Additionally, during training, random masking is applied to the generated spectrogram features in either the frequency or time dimension.

\textbf{Data Normalization}: Conventionally, entire spectrograms are normalized. In contrast, this study normalizes each row of the spectrogram individually. Considering that each row in the spectrogram represents information within a specific frequency range and different rows represent different frequency bands, normalizing rows aids in identifying frequency variations within each row. Moreover, as respiratory sound features predominantly occur in the low-frequency part of the spectrogram, normalizing columns would potentially mask minor anomalies present in the high-frequency part. Therefore, this study adopts row-wise normalization for respiratory sound feature normalization.

\subsubsection{Parameter setting }

During preprocessing, all audio is resampled to 10 kHz and fixed to a duration of 8 seconds. Parameters used in the Short-Time Fourier Transform (STFT) for converting raw audio to spectrograms are set as follows: the number of FFT points $n\_fft$ = 1024, window length in points $win\_len$ = 1000, hop length in points $hop\_len$ = 128. The lowest and highest cutoff frequencies in spectrogram features are set to $f\_min$ = 32.7 and $f\_max$ = 3000 respectively, with a uniform filter bank number of 84. Each spectrogram generated contains approximately 625 frames per sample, with a frame length of 20 frames per group and a 5-frame overlap between groups. These grouped spectrograms are then fed into the  GFE module for encoding.

In the training process, we use the same  optimizers and learning rates for the deep clustering module and contrastive learning module, the optimizer used the  Adam with an initial learning rate of 0.01 . The learning rate is decayed by a factor of 0.33 every 150 epochs to ensure training stability over the course of 600 epochs.

\begin{figure*}[htbp]
\centering
\begin{minipage}[t]{0.8\textwidth}
  \centering
  \includegraphics[width=\linewidth]{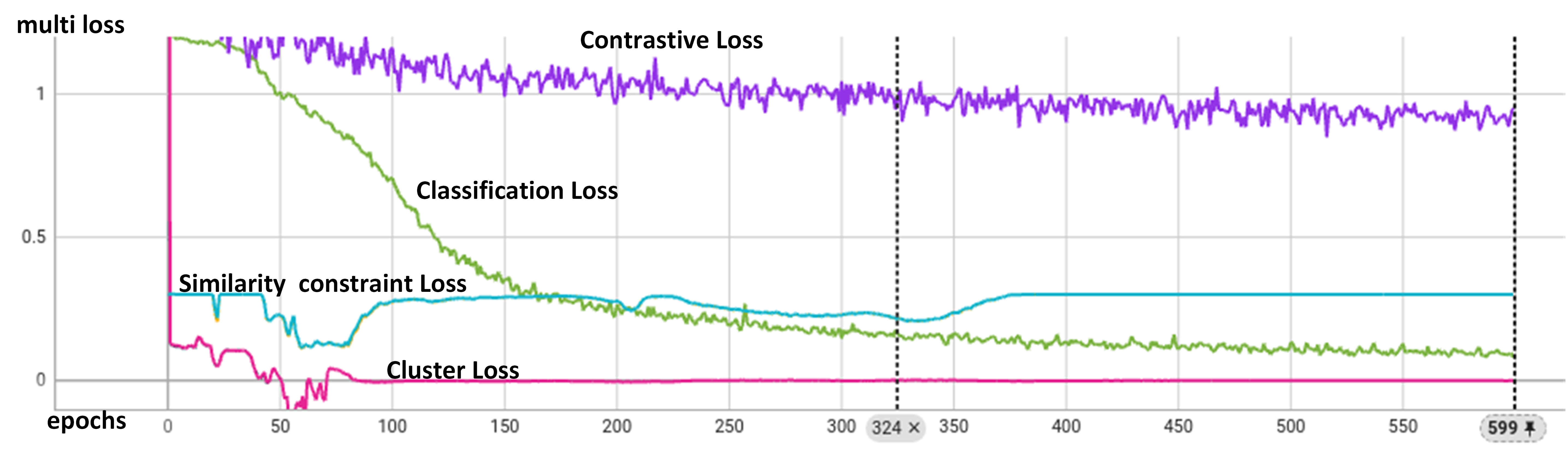}
  % \caption{Top subplot representing each loss item of Multiloss.}
  % \label{fig:sub1}
\end{minipage}
\quad % Add some horizontal space between subfigures
\begin{minipage}[t]{0.8\textwidth}
\centering
\includegraphics[width=\linewidth]{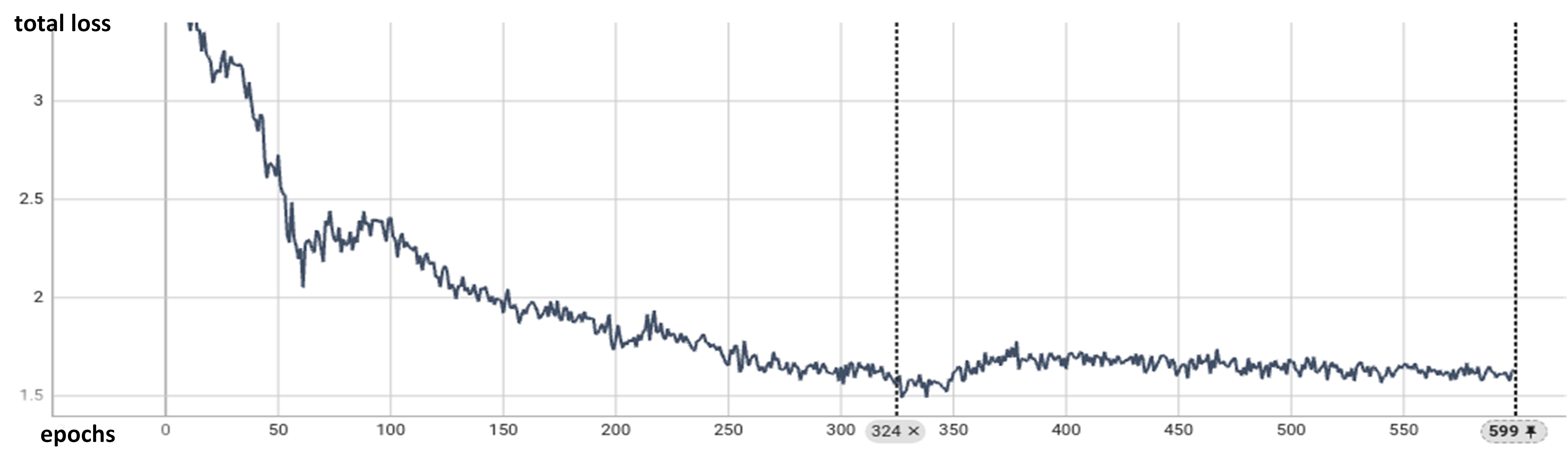}
  % \caption{Bottom subplot showing the total loss.}
  % \label{fig:sub2}
\end{minipage}
\caption{The upper show the Multiloss and bottom  show the total loss.}
\label{Fig.6}   %\label{Fig.5}
\end{figure*}

As show in {Fig.6}, the upper subfigure  demonstrates the steady reduction in contrast loss and classification loss over the course of training, indicating that the contrast learning as well as the classification head learns the respiratory sounds features better. The values of similarity loss and clustering loss are stable in an interval, which shows that the clustering module does play the role of auxiliary constraints between different cluster features.The bottom subfigure  demonstrates the steady reduction in total loss.

\subsection{Comparative study}

\subsubsection{Comparing with other method}
To validate the effectiveness and superiority of the proposed method, we utilized the official 6-4 validation split and compared the classification performance of various current methods on the ICBHI2017 dataset, as shown in {Table2}.

For the Specificity metric, we bolded scores above 80. Among these works, models achieving high scores without the influence of pre-trained weights employed an attention mechanism-based approach for feature fusion. This suggests that, for the entire spectrogram, selective feature extraction is more beneficial than extracting all information, as it enhances the model's ability to distinguish between normal and abnormal respiratory sounds. 

Regarding the Sensitivity metric, we bolded values above 41. The patch-based method, without pre-trained weights, achieved a sensitivity score of around 30. However, with pre-trained weights\cite{song2023patch}, \cite{bae2023patch}, the model's sensitivity significantly increased, indicating that pre-trained weights on large-scale datasets help capture subtle feature differences. Additionally, in the \cite{xu2021arsc} literature, although they did not use a patch-based approach, they employed a 7x7  convolution kernel, which approximates patch extraction. This further demonstrates that extracting local features from the spectrogram aids in capturing differences between abnormal respiratory sounds.

In terms of the score, our method, without pre-trained weights, outperforms the\cite{xu2021arsc} method by nearly 7 points and the \cite{bae2023patch} method by about 13 points. Compared to models with available pre-trained weights, our method outperforms \cite{bae2023patch} by nearly 1 point. Regardless of the availability of pre-trained weights, our model achieves the best balance between specificity and sensitivity.

Regarding model size, our architecture is entirely self-designed, with model parameters amounting to 38MB. In contrast, the\cite{bae2023patch} method utilizes a Transformer-based network architecture, with a size exceeding 380MB. The network architecture we designed strikes the best current balance between parameter size and performance, making it suitable for deployment on memory- and compute-constrained mobile devices.

\begin{table*}[htbp]
\flushleft
\captionsetup{justification=raggedright}
\caption{ Performance and size comparison with state-of-the-art models, where \textbf{IN, AS} represent used the pre-training weights on ImageNet and AudioSet respectively, and \textbf{'-'} represents not used any pre-training weights, spliting the dataset with the offical method. }

\setlength{\tabcolsep}{4mm}{
\begin{tabular}{lccccccccc} % 控制表格的格式
\toprule
% \multirow{1}{*} Model Number   \\
%\cline{2-6}  % 这部分是画一条横线在2-6 排之间
Method  &  Architecture/ Pretrained  &  $S_p$   &  $S_e$ &  $score$  &Size  \\
\midrule
 4 Class    &                     &                     &                        &                  &           \\
Path-level\cite{song2023patch}        & CNN6 /IN            &     68.59 $\%$      & \textbf{48.25} $\%$    &  58.42 $\%$      &  \textbf{20}MB+ \\
RespirNet\cite{gairola2021respirenet} & Resnet34/IN         &     72.30 $\%$      &   40.10    $\%$        &  56.20 $\%$      &    83MB+      \\
Domain transfor\cite{wang2022domain}  &  Resnet-st/IN       &   70.40 $\%$        &   40.20    $\%$        &   55.30 $\%$     &    50MB+      \\
Late fusion\cite{pham2022ensemble}    &  Ince-v3+VGG14/IN   & \textbf{85.60} $\%$ &   30.00    $\%$        &  57.30 $\%$      &    610MB+      \\
Co-tuing\cite{nguyen2022lung}         &  Resnet50/IN        &   79.34 $\%$        &   37.24    $\%$        &  58.29 $\%$      &    97MB+      \\               
SCL\cite{moummad2023pretraining}      & CNN6/AS             &   76.93 $\%$        &   39.15    $\%$        &  58.04 $\%$      &    20MB+      \\
Patch mix\cite{bae2023patch}          & AST / IN+AS         &   \textbf{81.66} $\%$        &   \textbf{43.07}    $\%$        &\textbf{62.37} $\%$ &   \textbf{380}MB+ \\
OFGST   \cite{wang2024ofgst}          &  Swin Transformer/ IN&  71.56 $\%$        &   40.53    $\%$        &  56.05 $\%$ &   410 MB+ \\
Stethoscope \cite{kim2024stethoscope} & AST/ IN+AS          &   79.87 $\%$        &   \textbf{43.55}    $\%$        &  61.71 $\%$ &   380MB+ \\
RepAug\cite{kim2024repaugment}        & AST/ IN+AS          &   \textbf{82.47} $\%$        &   40.55    $\%$        &  61.51 $\%$ &   380MB+ \\
Lungbrn \cite{ma2019lungbrn}          & bi-Resent/-         &    69.20 $\%$       &   31.12    $\%$        &  50.16 $\%$      &    80MB+      \\
SE+SA\cite{yang2020adventitious}      &   Resnet18/-        &    \textbf{81.25} $\%$       &   17.84    $\%$        &  49.55 $\%$      &    44MB+      \\ 
LungRN+NL \cite{ma2020lungrn+}        &    Resnet-NL/-      &    63.20 $\%$       &   41.32    $\%$        &   52.26 $\%$     &    40MB+      \\
Cnn+MOE\cite{pham2021cnn}             &    C-DNN/ -         &    72.40 $\%$       &   21.50    $\%$        &   47.00 $\%$     &    30MB+      \\
LungAttn \cite{li2021lungattn}        &   ResNet-Atten/-    &    71.44 $\%$       &   36.36    $\%$        &   53.90 $\%$     &    60MB+      \\
ARSC-net \cite{xu2021arsc}            &   bi-Resnet-Att/-   &     67.13 $\%$      &   \textbf{46.38} $\%$  &  \textbf{56.76} $\%$      &    80MB+      \\  
Prototype\cite{ren2022prototype}      &   Cnn8-pt/-         &     72.96 $\%$      &   27.78    $\%$        &   50.37 $\%$     &    30MB+      \\
Adversarial \cite{chang2022example}   &   Cnn8-dilated/ -   &   69.92 $\%$        &   35.85    $\%$        &  52.89 $\%$      &  30MB+      \\
Patch mix\cite{bae2023patch}          & ASTransformer/ -    &   72.61 $\%$        &   30.58    $\%$        &  \ 49.60 $\%$    &   \textbf{380}MB+ \\
\textbf{Ours(cos)}                    & CycleGuardian/-     & \textbf{88.92} $\%$ &   33.33    $\%$        &  \textbf{61.12} $\%$ &    \textbf{27}MB     \\
\textbf{Ours(soft cos)}               & CycleGuardian/-     & \textbf{82.06} $\%$ &   \textbf{44.47}    $\%$        &  \textbf{63.26} $\%$ &    \textbf{38}MB     \\

\midrule
2 Class                               &                       &                      &                        &              &         \\
Co-tuing\cite{nguyen2022lung}         & Resnet50 /IN          &  \textbf{79.34} $\%$ &   50.04 $\%$  &     64.74 $\%$        &    97MB+      \\
Patch mix\cite{bae2023patch}          & ASTransformer/ IN +AS & \textbf{81.66} $\%$  &  \textbf{55.77} $\%$  &   \textbf{68.71} $\%$ &    380MB+      \\
Cnn+MOE\cite{pham2021cnn}             &   C-DNN/-             &     72.40 $\%$       &   37.50 $\%$         & 54.10 $\%$           &    30MB+      \\                 
\textbf{Ours}                         &   CycleGuardian/-     &   73.19 $\%$         &  \textbf{59.57} $\%$  & \textbf{66.38} $\%$ &  \textbf{27}MB      \\
\textbf{Ours(soft cos)}               & CycleGuardian/-     & \textbf{71.93} $\%$ &   \textbf{62.77} $\%$   &  \textbf{67.35} $\%$  &    \textbf{38}MB     \\
\bottomrule
\end{tabular} 
    }
\flushleft
  \label{Table2}
\end{table*}

To analyze the bottlenecks of our current model, we computed the confusion matrix based on the official validation split, as illustrated in {Fig.7}. Using this confusion matrix, we calculated the precision, recall, and F1-score for each class using the following formulas $\text{Precision} = \frac{TP}{TP + FP}, \quad \text{Recall} = \frac{TP}{TP + FN}, \quad \text{F1-score} = 2 \times \frac{\text{Precision} \times \text{Recall}}{\text{Precision} + \text{Recall}}$. The results are shown in {Table3}:

\begin{figure}[htbp]
\begin{center}
\includegraphics[width=0.4\textwidth]{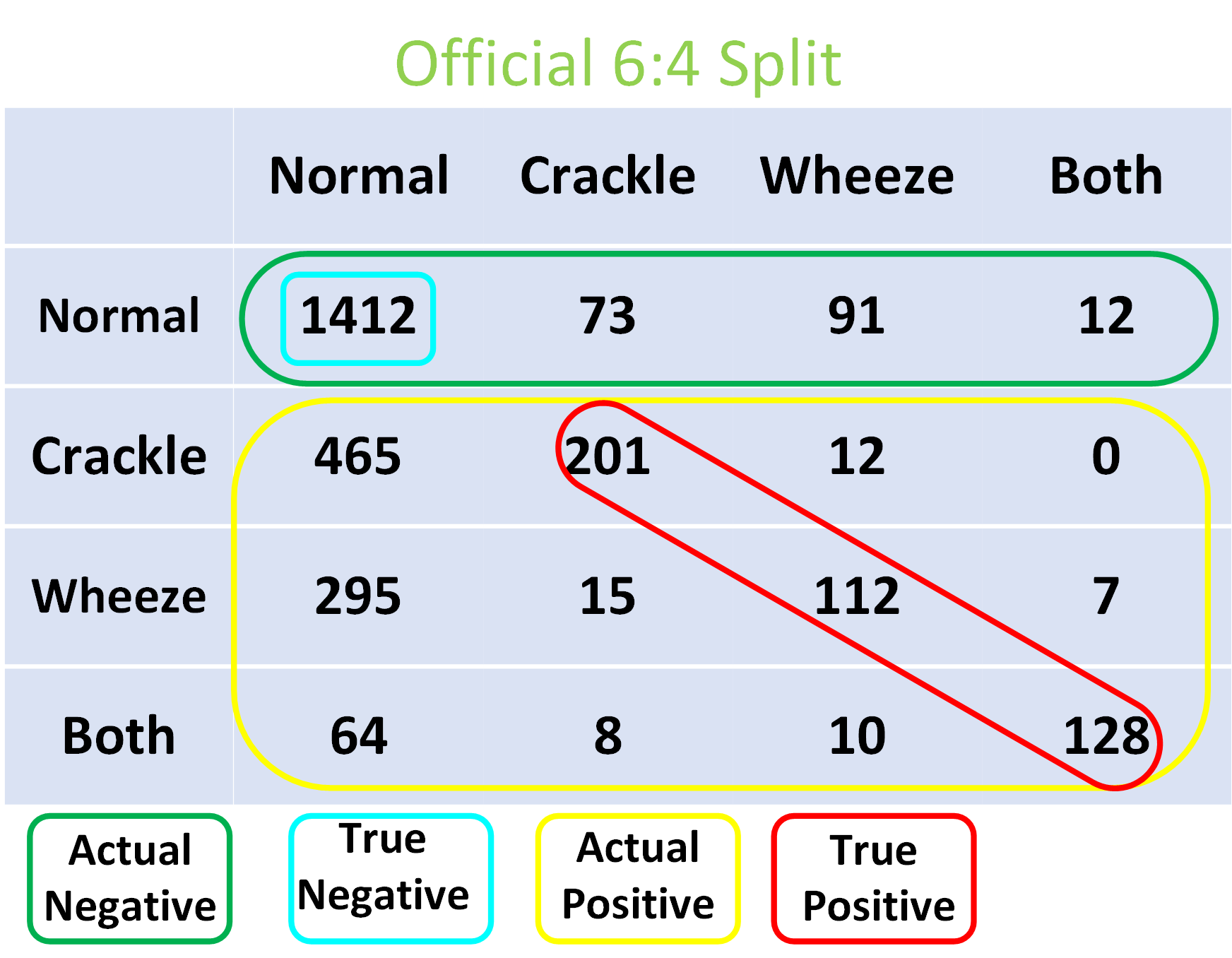}    
\end{center}
\caption{Confusion matrices corresponding to the official 6-4 divisions }
\label{Fig.7}
\end{figure}

\begin{table}[htbp]
  \flushleft
  \captionsetup{justification=raggedright}
  \caption{Model performances on four class}
  \setlength{\tabcolsep}{2mm}{
  \begin{tabular}{lccc} % 控制表格的格式
  \toprule
  % \multirow{1}{*} Model Number   \\
  % \cline{2-6}  % 这部分是画一条横线在2-6 排之间
   Class &  Precision  & Recall &    F1-score    \\
  \midrule
    Normal  &  63.14 $\%$    &  88.92 $\%$   &  73.87 $\%$       \\
    Crackle  &  67.67 $\%$   &    29.61 $\%$ &  41.18 $\%$       \\
    Wheeze  &  49.78 $\%$    &   26.10 $\%$ &  34.23 $\%$       \\
    Both  &   87.07 $\%$     &   60.95 $\%$ &  71.70 $\%$       \\
  \bottomrule
  \end{tabular} 
      }
  \flushleft
    \label{Table3}
  \end{table}

From the {Table3}, it can be observed that despite the normal class having the most samples in the dataset (2063), and the "both" class having the fewest samples (363), the model achieves higher precision in the "both" category than in the normal category. The F1-score for the "both" category is also close to that of the normal category, indicating that the model performs well in detecting symptoms in the "both" class. However, the recall rates for the crackle and wheeze abnormal categories are notably low. To further compare the confusion between abnormal and normal classes, we calculated the true positives (TP), false positives (FP), and false negatives (FN) for each abnormal category relative to the normal category, with the results presented {Table4}.

\begin{table}[htbp]
  \flushleft
  \captionsetup{justification=raggedright}
  \caption{ Abnormal category confused with the normal category }
  
  \setlength{\tabcolsep}{2mm}{
  \begin{tabular}{lccc} % 控制表格的格式
  \toprule
  % \multirow{1}{*} Model Number   \\
  % \cline{2-6}  % 这部分是画一条横线在2-6 排之间
   Class pair & TP & FP &    FN    \\
  \midrule
   Crackle vs. Normal  &    201    &  73  &  465       \\
   Wheeze vs. Normal  &  112      & 91 &  295       \\
   Both vs.  Normal  &  128     & 12 &  64        \\
  \bottomrule
  \end{tabular} 
      }
  \flushleft
    \label{Table4}
  \end{table}

{Table4} indicates that the model is less likely to incorrectly classify normal samples as abnormal, corresponding to a lower False Positive rate. However, it is more prone to misclassifying abnormal samples as normal, corresponding to a higher False Negative rate. This leads to the model’s lower recall rate for crackle and wheeze abnormal respiratory sounds.

\subsubsection{Different  Patients }

To evaluate the model's generalization performance across different patients, we conducted a five-fold cross-validation, where each fold used a distinct set of patients for validation. Each patient's respiratory sound segments appeared exclusively in either the training or validation sets, maximizing the assessment of the model's generalization across different patients. {Table5} shows the proportion  of respiratory sound samples per class in each fold, and {Fig.8} presents heatmaps of the confusion matrices for each fold.

\begin{table}[htbp]
  \flushleft
  \captionsetup{justification=raggedright}
  \caption{The ratio of respiratory samples per class in five folds with different patients}
  
  \setlength{\tabcolsep}{2mm}{
  \begin{tabular}{lccccc} % 控制表格的格式
  \toprule
  % \multirow{1}{*} Model Number   \\
  % \cline{2-6}  % 这部分是画一条横线在2-6 排之间
   Fold &  Normal  & Crackle & Wheeze & Both & Numbers     \\
  \midrule
0 & 0.60/0.45  &  0.19/0.34   &  0.16/0.10   &  0.05/0.10  & 3433/3455   \\
1 & 0.51/0.65  &  0.28/0.12   &  0.12/0.15   & 0.07/0.08   & 6194/704    \\
2 & 0.51/0.75  &  0.28/0.20   &  0.14/0.04   & 0.08/0.01   & 6362/536    \\
3 & 0.52/0.58  &  0.29/0.11   & 0.11/0.27    & 0.08/0.04   & 6139/759     \\
4 & 0.52/0.54  &  0.27/0.27   & 0.12/0.14    &  0.08/0.04  & 5454/1444    \\
  \bottomrule
  \end{tabular} 
      }
  \flushleft
    \label{Table5}
  \end{table}
  
\begin{figure*}[htbp]
\begin{center}
\includegraphics[width=0.9\textwidth]{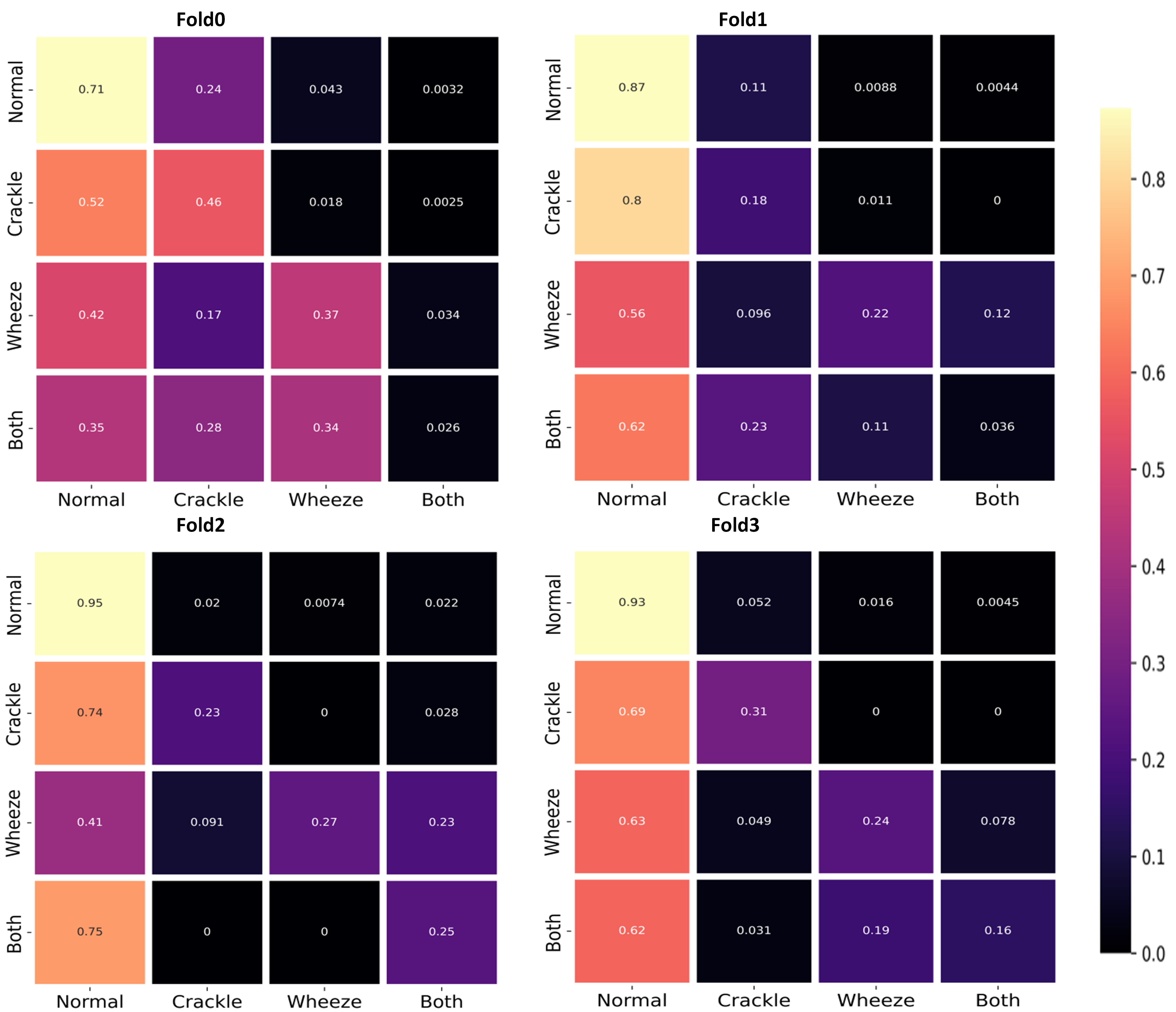}    
\end{center}
\caption{ Model  generalization performance test,  confusion matrices with heatmaps, from fold0 to fold4 patient data in each fold is different from other folds. }
\label{Fig.8}
\end{figure*}

According to {Fig.8}, in Fold 0, the model's performance in recognizing crackle showed no significant change, but wheeze and crackle were often confused, and the model completely failed to identify the "both" category, instead classifying it as either crackle or wheeze. In Fold 1, the recognition performance for all three abnormal symptoms declined, with increased confusion between wheeze and both, as well as between both and crackle. Similar patterns were observed in Folds 2 and 3, where the recognition performance for the three abnormal symptoms decreased, and the confusion between wheeze and both increased. Conversely, in Fold 4, the model maintained stable recognition performance for crackle and wheeze, but there was significant confusion between both and wheeze.

Overall, the model demonstrated strong recognition ability for normal samples, likely due to the large number of normal samples in the dataset. However, the recognition of the "both" category exhibited significant fluctuations, which can be attributed to the small number of both samples. When the distribution of these samples changes, it greatly impacts the model's performance. The significant variations in recognizing abnormal respiratory sounds across different patients suggest that a major factor contributing to the model's poor generalization performance is the insufficient number of abnormal samples in the dataset.

To understand the generalization performance of other models and methods in this field, we compared their performance using an 8-2 split, as shown in {Table6}. Additionally, the confusion matrix of our model under the 8-2 split is presented in {Fig.9}, since the 8-2 division corresponds to fold4 in the 5-fold division, we put it together.

\begin{table*}[htbp]
  % \flushleft
  \captionsetup{justification=raggedright,singlelinecheck=false}
  \caption{Comparison of other methods under the 8-2 split approach}
  \setlength{\tabcolsep}{3mm}
  {
  \begin{tabular}{lccccc} % 控制表格的格式
  \toprule
  % \multirow{1}{*} Model Number   \\
  % \cline{2-6}  % 这部分是画一条横线在2-6 排之间
  Method &  Architecture/ Pretrained   &   $S_p$    &  $S_e$ &  $score$   & Size  \\
  \midrule
  RespirNet \cite{gairola2021respirenet}     &  Resnet34/IN     &     78.80 $\%$   &     53.62 $\%$    &  66.20 $\%$    &    83MB+      \\   
  Patient tune\cite{acharya2020deep}         &   CNN-RNN / IN    &   84.10 $\%$     &     48.60 $\%$    &  66.30 $\%$   &    200MB+      \\
  LungRN+NL \cite{ma2020lungrn+}             &   Resnet-NL/-      &   64.71 $\%$   &\textbf{63.70} $\%$  &   64.20 $\%$  &    40MB+      \\
  LungAttn \cite{li2021lungattn}             &   ResNet-Atten/- &   80.45 $\%$     &     54.69 $\%$    & \textbf{67.57} $\%$ &    40MB+      \\
  Noise mask\cite{kochetov2018noise}         &   Noise rnn /-   &   73.0 $\%$      &     58.40 $\%$    &  65.70 $\%$ &    110MB+      \\
  \textbf{Ours}                              &   CycleGuardian&\textbf{91.20} $\%$ &    37.66 $\%$    &  64.44 $\%$ &    27MB      \\
  \bottomrule
  \end{tabular} 
   }
  %\flushleft
  \label{Table6}
  \end{table*}
  
  As indicated in {Table6}, when the overall data distribution in the training and test sets changes, the performance of existing models shows substantial variability. Although the works of \cite{gairola2021respirenet} \cite{ma2020lungrn+} \cite{li2021lungattn} \cite{kochetov2018noise} report lower sensitivity scores under the official 6-4 split, they achieve better sensitivity scores under the 80-20 split. Currently, no model or method with robust generalization performance has consistently performed well across different split configurations.

\begin{figure*}[htbp]
\begin{center}
\includegraphics[width=0.8\textwidth]{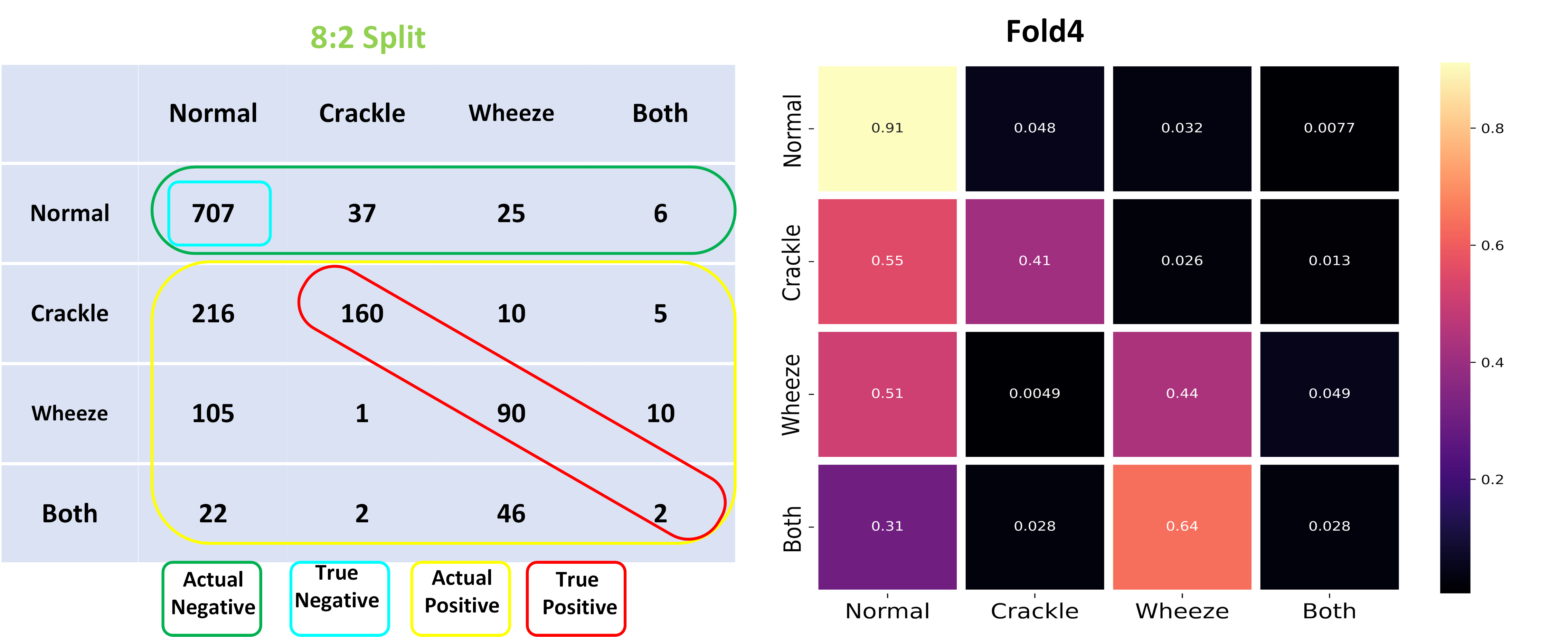}    
\end{center}
\caption{Confusion matrices  with heatmaps corresponding to the  8-2 divisions }
\label{Fig.9}
\end{figure*}

{Fig.9}'s confusion matrix shows that under the 8-2 split, the features learned for the "both" and "wheeze" abnormal types are significantly confused, whereas the model can better distinguish these two types under the 6-4 split. This further suggests that the lack of abnormal samples leads to insufficient feature representation learning for abnormal samples, resulting in significant performance variation when the dataset distribution shifts. On the other hand, in both the official and 8-2 splits, the model consistently distinguishes normal respiratory sounds from abnormal ones, but the highest confusion occurs between abnormal respiratory sounds and normal ones. We hypothesize that this is due to shared components between normal and abnormal respiratory sounds, which include noise and normal respiratory elements. The extent to which these shared components contribute to the confusion is further analyzed in the noise ablation experiment.

\subsubsection{Impact of stethoscope devices }

The ICBHI dataset was constructed using four different types of stethoscopes: Meditron, LittC2SE, Litt3200, and AKGC417L. To further assess the model's generalization performance across different stethoscopes, we conducted tests where the data corresponding to a specific stethoscope A was excluded from the training set when evaluating the generalization on that stethoscope A.

\begin{table}[htbp]
  \flushleft
  \captionsetup{justification=raggedright}
  \caption{The ratio of samples for each class with four stethoscopes}
  
  \setlength{\tabcolsep}{1mm}{
  \begin{tabular}{lccccc} % 控制表格的格式
  \toprule
  % \multirow{1}{*} Model Number   \\
  % \cline{2-6}  % 这部分是画一条横线在2-6 排之间
  Stethoscope &  Normal  & Crackle & Wheeze & Both & Numbers     \\
  \midrule
  Meditron & 0.70/0.64 &  0.15/0.18  & 0.10/0.10   &  0.05/0.08  & 1048/494     \\
  LittC2SE & 0.59/-    &  0.13/-     & 0.21/-      &  0.07/-     & 594/0       \\
  Litt3200 & 0.27/0.69 &  0.10/0.06  & 0.39/0.20   & 0.24/0.04   &  49/474      \\
  AkGC417L & 0.39/0.50 &  0.39/0.29  & 0.10/0.13   & 0.12/0.08   & 2571/1905     \\
  \bottomrule
  \end{tabular} 
      }
  \flushleft
    \label{Table7}
  \end{table}

{Table7} presents the proportion of samples for each class from each device in the training and validation sets under the official split. Notably, although the test set includes samples from all four respiratory sound classes, it does not contain data from the LittC2SE device. Therefore, we only evaluated the generalization performance on the other three devices.

\begin{figure*}[htbp]
\begin{center}
\includegraphics[width=0.8\textwidth]{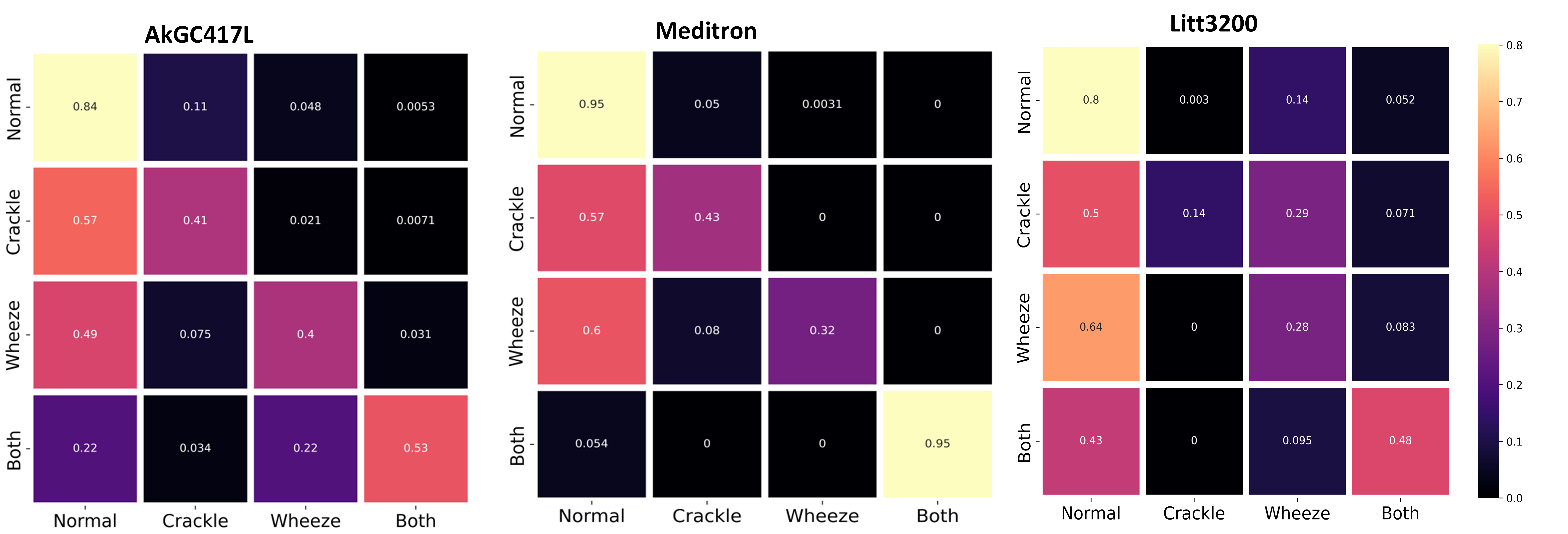}    
\end{center}
\caption{Generalization test on different stethoscope devices.}
\label{Fig.10}
\end{figure*}

As shown in the {Fig.10}, the current model demonstrates the best generalization performance on the Meditron device, particularly achieving the same level of accuracy in identifying the "both" category as the "normal" category, despite the significant difference in sample sizes between these two classes. We believe that, in addition to the fact that Meditron uses a digital electret microphone as its auscultation sensor, its multiple filtering options for isolating specific frequencies work well with the proposed grouped feature extraction method, leading to higher recognition accuracy.

The model's generalization performance on the AKGC417L device remains relatively stable, with no significant fluctuations. In contrast, the performance on the Littmann 3200 device shows a decline, especially with increased confusion between the crackle and wheeze classes. We attribute this to the differences in the data collected by this device's sensor compared to the others, particularly for crackle sounds. The Littmann 3200 uses a piezoelectric sensor, which provides sound amplification and noise reduction, making it useful in noisy environments or when detecting faint sounds. Additionally, similar to the Meditron device, it offers multiple filtering options to focus on specific frequency ranges.

Overall, the tests on three different stethoscopes demonstrate that the proposed model exhibits good generalization performance across different stethoscope devices.

\subsubsection{Impact of group frames }
Each sample spectrogram consists of 626 frames. To evaluate the impact of the number of frames per group, experiments were conducted using groups composed of 2-25 frames each. In {Fig.11}, "frames" represents the number of frames per group, and "groups" indicates the total number of groups under each case. The grouping process utilized floor division, discarding any remaining frames unable to form a complete group.

\begin{figure}[htbp]
\begin{center}
\includegraphics[width=0.40\textwidth]{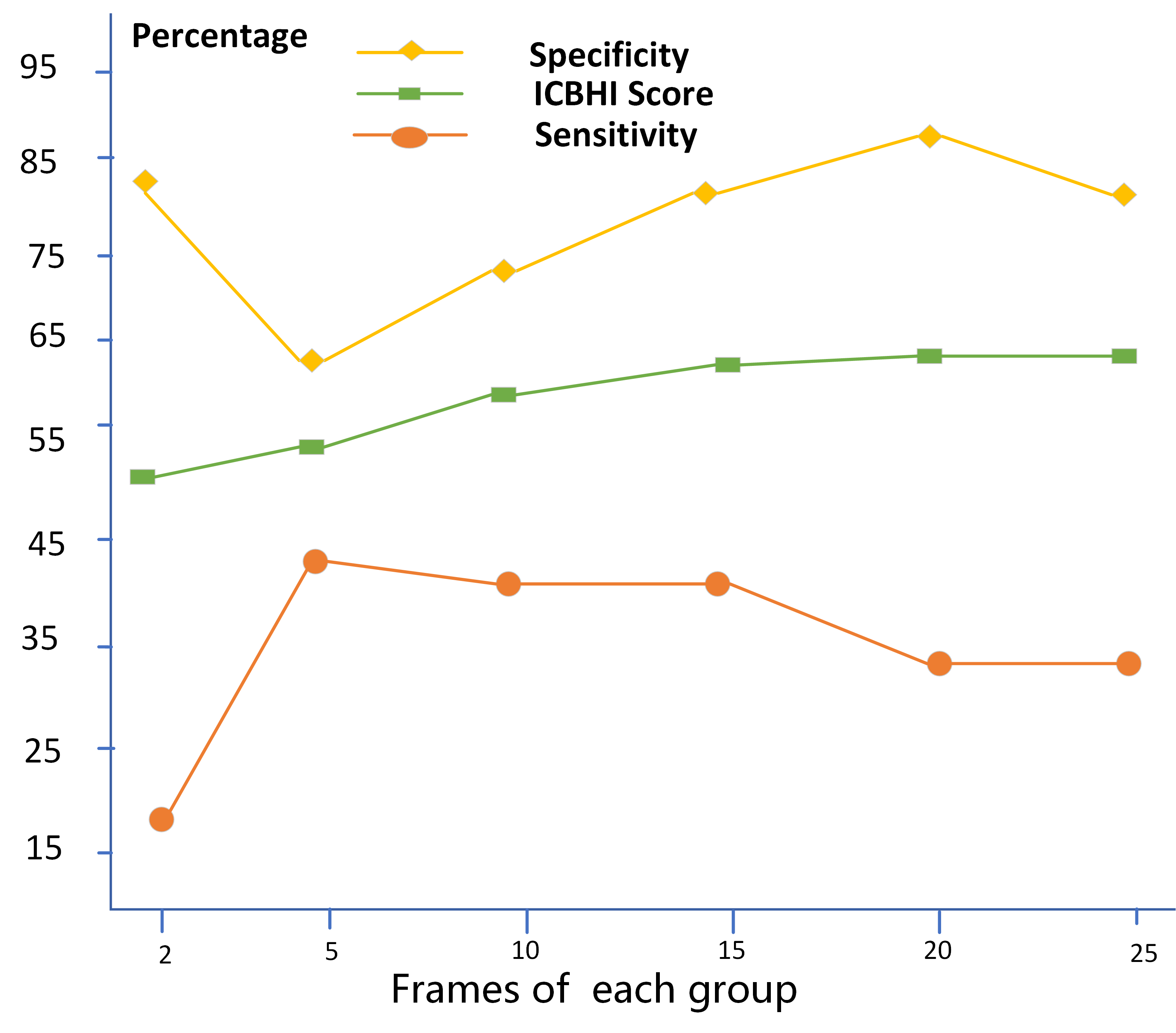}    
\end{center}
\caption{Effect of the number of frames in each group on specificity and sensitivity, respectively.}
\label{Fig.11}
\end{figure}

Analysis from {Fig.11} reveals that when the number of frames per group ranges between 5 and 15, sensitivity scores are generally higher, indicating better ability of the model to discern abnormal respiratory sound features. However, when the number of frames per group is only 2, there is a significant decrease in sensitivity score, possibly due to too few frames per group leading to excessive grouping and decreased performance of the deep clustering module. On the other hand, an overall increasing trend in specificity scores is observed with an increase in the number of frames per group, suggesting that moderately increasing the number of frames per group facilitates distinguishing between normal and abnormal respiratory sound features. Similarly, beyond 20 frames per group, there is no further advantage in distinguishing between normal and abnormal features. Ultimately, obtaining 20 frames per group yields the best balance between specificity and sensitivity scores.

\subsection{Ablation  study }

\subsubsection{Data augmentations }
To validate the impact of different data augmentations, our study conducted experiments from two aspects: one involved altering the original audio through speed adjustment or shifting, while the other involved applying random masking to spectrograms. Specifically, we use adding noise, shifting, stretching, and VTLP with 50 $\%$ probability in different batches, and approximately 30$\%$ of the temporal or frequency dimensions of spectrograms were masked. Samples within the same batch underwent identical data augmentation. {Table8} demonstrates that data augmentation at the raw audio level increased specificity  from the original 76$\%$ to 80$\%$,  with random masking on spectrograms specificity and sensitivity increased from the original 76$\%$ and 36$\%$ to 77$\%$ and 38$\%$, respectively. Ultimately, combining both types of data augmentation yielded the better results.

\begin{table}[htbp]
\flushleft
\captionsetup{justification=raggedright}
\caption{Impact of data augmentations on model performance}

\setlength{\tabcolsep}{3mm}{
\begin{tabular}{lccccccccc} % 控制表格的格式
\toprule
% \multirow{1}{*} Model Number   \\
% \cline{2-6}  % 这部分是画一条横线在2-6 排之间
 Data Augementation &  Sp  & Se &   Score      \\
\midrule
Base DEC  &   76.46 $\%$     &36.21 $\%$   &56.33 $\%$        \\
Audio Aug  &   80.56 $\%$    &34.36 $\%$   &57.46 $\%$      \\
Spectrogram Aug  &   77.42 $\%$    & 38.22 $\%$   & 57.82 $\%$    \\ 
Audio + Spec  Aug  &    79.59 $\%$  &  37.53 $\%$  & 58.56 $\%$       \\
\bottomrule
\end{tabular} 
    }
\flushleft
  \label{Table8}
\end{table}

\subsubsection{Module functions}

Building upon data augmentation, we conducted ablation experiments to further validate the effectiveness of improved deep embeding cluster based on  the similarity constraint  and the contrastive learning module. As shown in {Table9}, comparing to the basic DEC module, the improvement in specificity from 76 $\%$ to 86$\%$ , this implies that the cosine similarity-constrained deep clustering module facilitates distinguishing features between normal and abnormal respiratory sounds. On the other hand, with the introduction of the contrastive learning module, increasing the  sensitivity from 36$\%$ to 39$\%$, this demonstrates the effectiveness of the contrastive learning with group mix module in distinguishing between three classes  of different abnormal features. Moreover, the highest score 61$\%$  was achieved when both constrained clustering and contrastive learning functionalities were simultaneously introduced.After replacing the cosine similarity-based module with a learnable soft cosine similarity-based module, the model's ability to distinguish between abnormal classes has been further enhanced.

\begin{table}[htbp]
  \flushleft
  \captionsetup{justification=raggedright}
  \caption{Effects of improved deep embedding  cluster,  contrastive learning with group mix, soft cos similarity on model performances. }
  \setlength{\tabcolsep}{2mm}{
  \begin{tabular}{lccccccccc} % 控制表格的格式
  \toprule
  % \multirow{1}{*} Model Number   \\
  % \cline{2-6}  % 这部分是画一条横线在2-6 排之间
   Basic Module &  Sp  & Se &   Score      \\
  \midrule
   Base  DEC 							&    76.46 $\%$     &   36.21  $\%$   &   56.33 $\%$         \\
   Imporved DEC (IDEC)   	&    86.84 $\%$   	&   32.08  $\%$   &   59.46 $\%$    \\ 
   Group Contrastive learning (GCL) 				  &   82.31  $\%$   &   39.33 $\%$  & 60.82 $\%$       \\
   IDEC + GCL (cos)   		&    88.92 $\%$     &   33.32  $\%$   &   61.12 $\%$    \\
   IDEC + GCL (soft cos)  &    82.06 $\%$     &   44.47  $\%$   &  63.26  $\%$   \\
  
  \bottomrule
  \end{tabular} 
      }
  \flushleft
    \label{Table9}
  \end{table}

\subsubsection{Noise ablation}

As previously discussed, the highest confusion for abnormal respiratory sound samples occurs with the normal respiratory sound class. We hypothesize that this confusion arises due to shared components between normal and abnormal respiratory sounds, specifically noise and normal respiratory elements. To investigate which component is the primary contributor to this confusion, we conducted a noise ablation experiment, with the results presented in {Fig.12}.

\begin{figure*}[htbp]
\begin{center}
\includegraphics[width=0.70\textwidth]{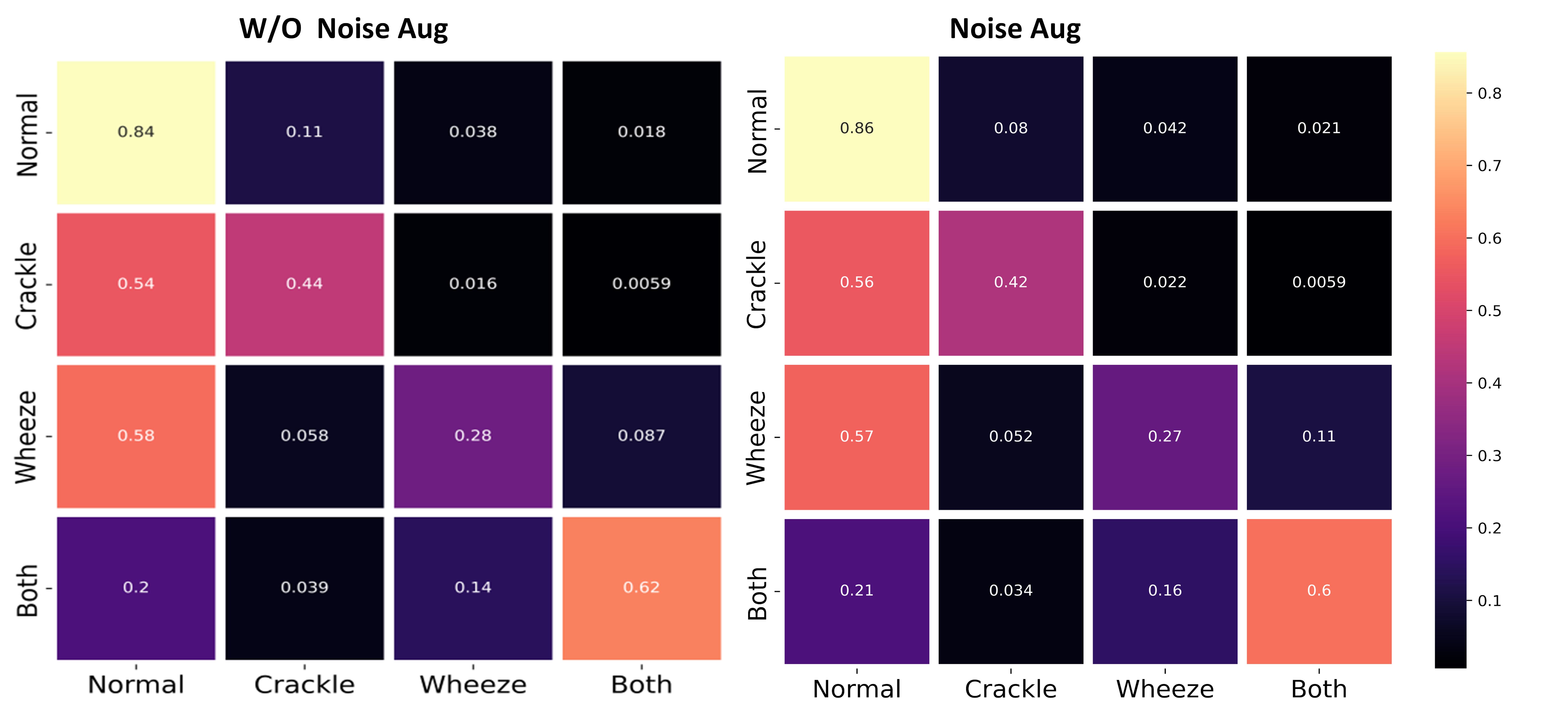}    
\end{center}
\caption{The confusion matrix on four class  without  and with  noise augmentation.}
\label{Fig.12}
\end{figure*}

From the {Fig.12}, it is evident that adding noise improves the model's ability to distinguish normal samples, but does not enhance the differentiation between abnormal and normal samples. Therefore, we conclude that noise is not the primary factor contributing to the confusion between abnormal and normal respiratory sounds.

This leaves the normal respiratory sound component as the remaining shared factor between normal and abnormal respiratory sounds. To further explore this, we analyzed the sample count per class across various time segments in the dataset, with the results as follows {Table10}:

\begin{table}[htbp]
  \flushleft
  \captionsetup{justification=raggedright}
  \caption{The number of samples per class on different time durations.}
  \setlength{\tabcolsep}{2mm}{
  \begin{tabular}{lcccc} % 控制表格的格式
  \toprule
  % \multirow{1}{*} Model Number   \\
  % \cline{2-6}  % 这部分是画一条横线在2-6 排之间
Duration Bin &  Normal  & Crackle & Wheeze & Both  \\
\midrule
0-100ms    & 0   &  0    & 0     &  0      \\
100-200ms  & 1   &  0    & 1     &  0      \\
200-300ms  & 6   &  0    & 1     &  0      \\
300-600ms  & 43  &  5    & 7  	 &  2       \\
600ms-1.5s & 628 &  113  & 112   &  14      \\
Above 1.5s & 2964&  1746 & 766   &  490     \\
  \bottomrule
  \end{tabular} 
      }
  \flushleft
    \label{Table10}
\end{table}

For crackle respiratory sounds, the abnormal duration typically ranges from 20-60 ms, with 95$\%$ of crackle cycles lasting over 1.5 seconds. Wheeze durations vary from 100 ms to 1 second, and for wheeze cycle-level annotations, 80$\%$ last over 1.5 seconds.

As shown in {Table10}, the annotated abnormal respiratory sound cycles in the current dataset contain a significant amount of normal respiratory sound components, especially within the annotated crackle segments, where normal components dominate. This explains why crackles are more likely to be misclassified as normal compared to wheezes. Upon analysis, we attribute this misclassification to the low proportion of abnormal components in these segments. The model’s classification results indicate that it follows a majority-rule principle, where the majority component determines the classification. Consequently, when an abnormal respiratory cycle predominantly contains normal components with only a minor portion of abnormal sounds, the model tends to misclassify it.

In light of this, we propose two areas for improvement in the current research on abnormal respiratory sound recognition.

\textbf{From the model perspective}:The challenge for current models is not merely recognizing difficult samples, but rather addressing the decision-making problem inherent in the task. Specifically, when a class contains elements of another class, the model must decide how to classify it. For example, given an abnormal respiratory sound that includes both normal and abnormal elements, how should the model decide to classify it as abnormal, especially when normal components are predominant (as in the case of crackles and wheezes). It is crucial to design a decision-making approach that accurately categorizes these as abnormal rather than normal.

\textbf{From the data perspective}: The existing dataset has two notable limitations: 1. The insufficient number of abnormal respiratory sound samples.  2. The high proportion of normal respiratory sound components within the annotated abnormal respiratory cycles. We recommend that clinicians, when annotating more abnormal respiratory sound samples, carefully select annotation intervals to ensure that abnormal components are in the majority and normal components are minimized within those intervals.

\subsection{Model  Deployment}
To facilitate the practical application of intelligent respiratory sound auscultation, recent efforts have introduced lightweight respiratory sound models, enabling deployment on wearable devices or mobile phone platforms \cite{chen2022diagnosis}. These studies focus on topics such as lightweight neural network design, network quantization, knowledge distillation, and pruning.

To achieve this goal, our designed network model was deployed on Android smartphones for experimentation. Initially, the respiratory  Android application was built using Android Studio on the host machine, and the trained model was converted into ONNX format using PyTorch. Subsequently, the corresponding ONNX runtime was installed on the Android device, providing the necessary runtime environment for executing ONNX models on mobile devices. Finally, the model was loaded and executed on the device by calling the interfaces provided by the ONNX runtime.

\begin{figure*}[htbp]
\begin{center}
\includegraphics[width=0.7\textwidth]{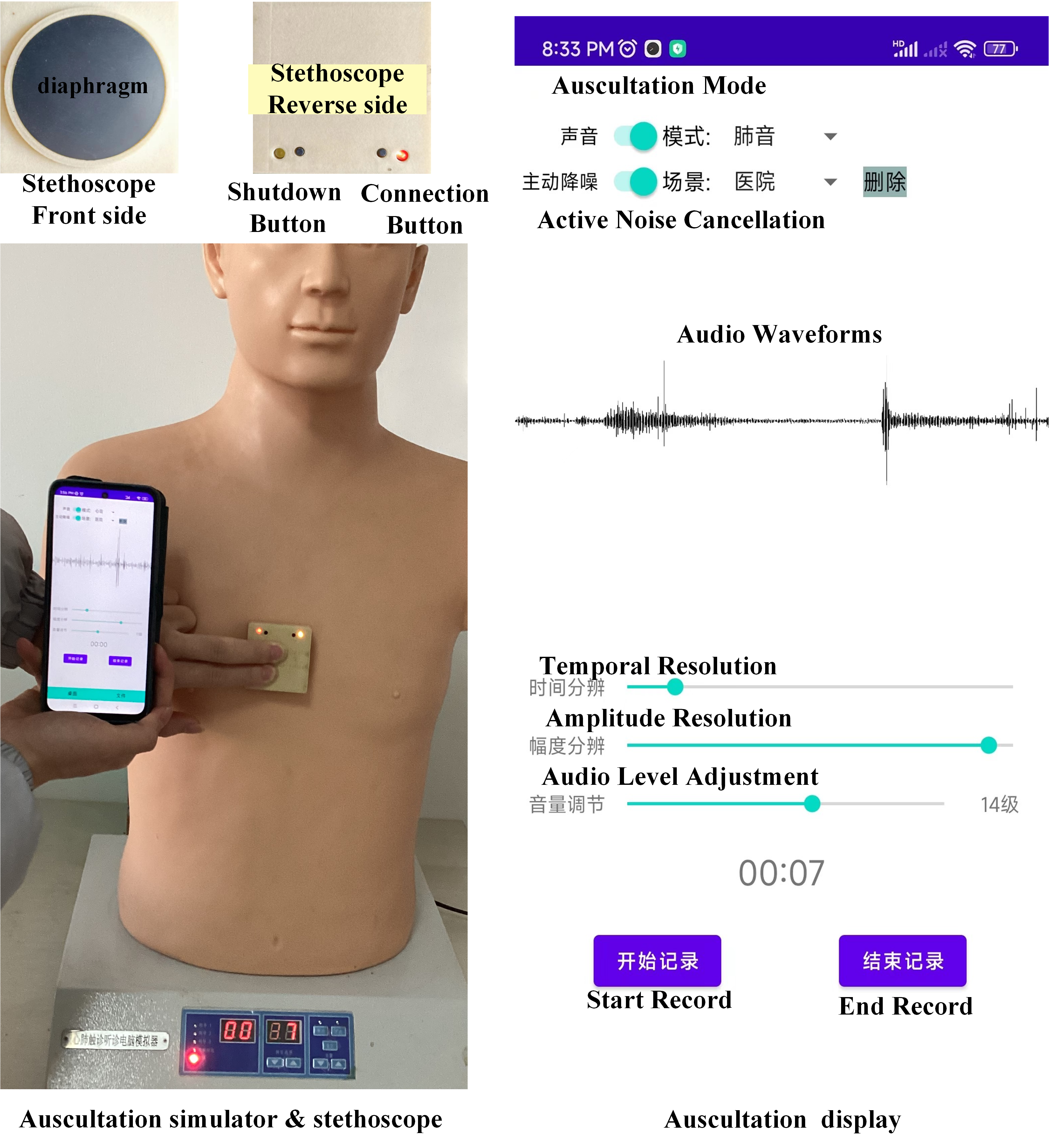}    
\end{center}
\caption{Respiratory sounds auscultation simulation: The left side shows the stethoscope used in this paper and the simulation of respiratory sounds auscultation, and the right side shows the real-time display of respiratory sounds waveforms on the Android client during the auscultation process.}
\label{Fig.13}
\end{figure*}

The diagram in {Fig.13} top left illustrates the stethoscope utilized in this study, employing a piezoelectric ceramic sensor as the pickup module. The hardware circuit primarily encompasses signal acquisition and analog-to-digital conversion, data buffering in the main control chip, Bluetooth transmission, and charging/discharging module. The left image of {Fig.13} bottom left depicts the process of acquiring respiratory sounds using the stethoscope. Initially, the stethoscope is paired and connected to a smartphone app via Bluetooth, followed by placement in an appropriate auscultation position on the chest. Then, the corresponding auscultation mode is selected in the app to commence respiratory sound collection. The right displays the respiratory sound signals collected during auscultation, which can be viewed in real-time on the smartphone and saved for further analysis.

{Fig.14} shows the prediction of the respiratory sounds collected by the network model deployed on Android, and shows the prediction probability of each category, taking the highest probability as the classification result. In addition, there is a choice of inference modes to find whether to use floating-point or integer modes for inference to  save computational resources. Comparing to the  transformer-based models that occupy approximately 380MB in size, our model occupies only around 38MB of space. This aspect is relatively favorable for mobile devices constrained by computational resources, memory, and power consumption. Through this demo, we have demonstrated the feasibility of constructing an intelligent respiratory sound auscultation system, providing a viable solution for detecting respiratory tract diseases through artificial intelligence.

\begin{figure*}[htbp]
\begin{center}
\includegraphics[width=0.7\textwidth]{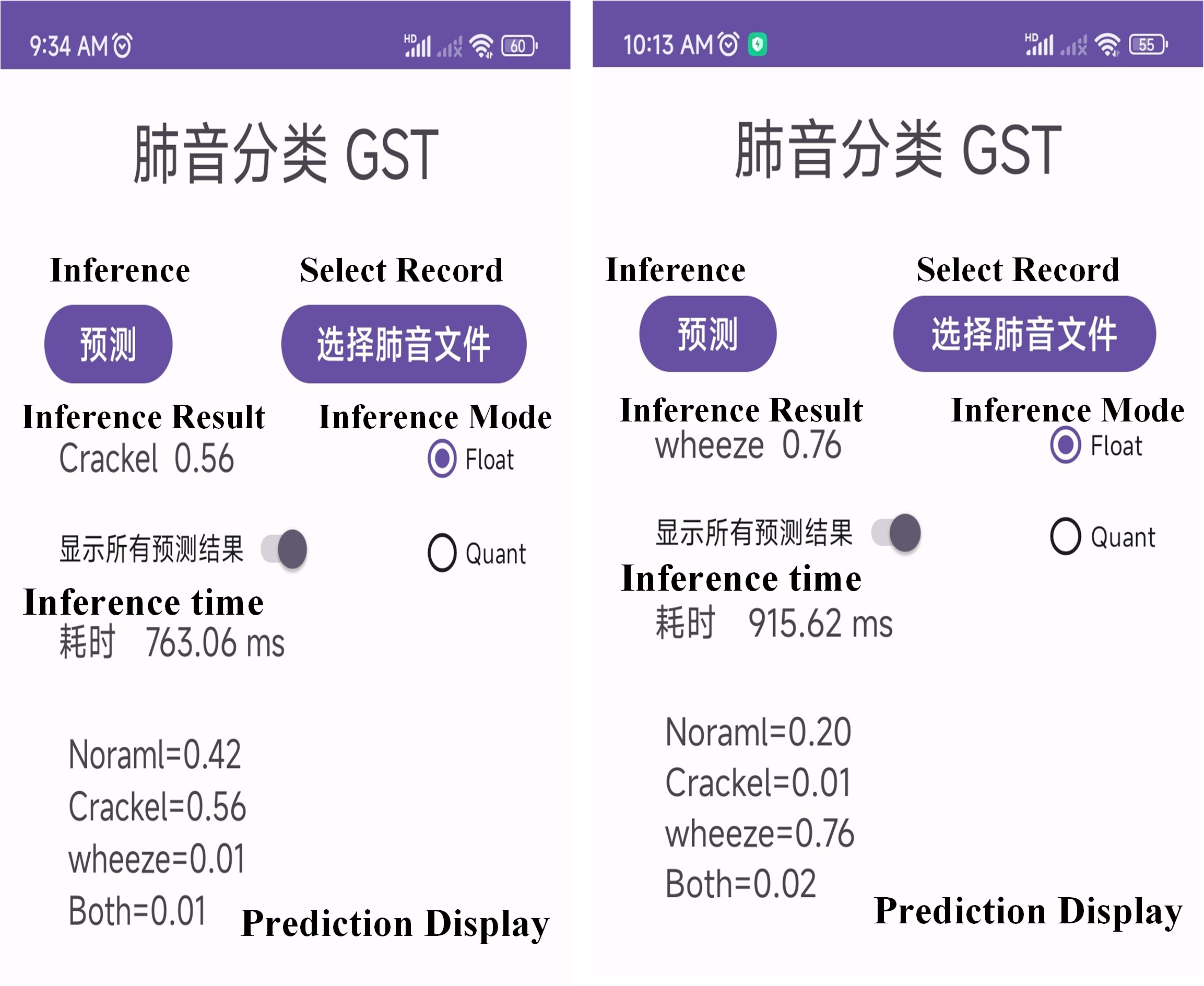}    
\end{center}
\caption{Automatic classification of recorded respiratory sounds by the model deployed on the Android SmartPhones.}
\label{Fig.14}
\end{figure*}

\section{Conclusion}
In this study, we present a novel automatic respiratory sound classification framework based on grouped spectrograms, incorporating Improved Deep Embedding Clustering (IDEC) and Contrastive Learning (CL) techniques, implemented in the lightweight network architecture, CycleGuardian. Compared to conventional approaches that directly classify spectrograms or apply patch-based methods, we explore a new feature encoding approach by grouping spectrogram features. We introduce an improved deep clustering and contrastive learning method within CycleGuardian. Experimental evaluation conducted on the ICBHI2017 dataset for multi-class respiratory sound classification, using the official 6-4 split, yields performance metrics of Sp: 82.06 $\%$, Se: 44.47$\%$, and Score: 63.26$\%$, with a network model size of 38M. Compared to various state-of-the-art approaches, without using pre-trained weights, our method achieves the best performances. Even, compared to the models utilizing pre-trained weights our  method still achieves the best balance between  performance and model size. Through ablation experiments, we demonstrate that deep clustering of grouped spectrograms enhances discrimination between normal and abnormal respiratory sounds, while contrastive learning based on grouped spectrograms improves discrimination among different types of abnormal respiratory sounds. Additionally, we tentatively deploy the network model for inference on Android mobile devices, showcasing a demo that illustrates the feasibility of constructing a complete intelligent respiratory sound auscultation system using deep learning techniques.

\begin{acknowledgements}
This work was supported in part by The Key Technologies Research and Development Program of China (2022YFC2404401)  and the Hainan Provincial Natural Science Foundation of China under Grant  81971692.
\end{acknowledgements}

\section{Declarations}

\textbf{Competing interests}
The authors declare that they have no known competing financial interests or personal relationships that could haveappeared to influence the work reported in this paper

\textbf{Ethics approval} 
This work adheres to the guidelines and ethical considerations outlined in the privacy policies of each social media platform involved in our study. We acknowledge the public nature and free accessibility of content posted on social network platforms, which are not password-protected and have thousands of active users. All analyses are
conducted using publicly available data, and we do not attempt to track
users across different platforms. In data preprocessing, we anonymized all user names or account IDs. Our data sets do not contain any information about individuals, and we have taken measures to ensure that
our results do not disclose the identity of any specific account

\textbf{Data availability}
The data set that support the findings of this study are
available via following: \url{https://bhichallenge.med.auth.gr/ICBHI_2017_Challenge}

\textbf{Code availability} 
Once our work has been accepted, we will open source our code.

\textbf{Author contribution}
Yun Chu: Methodology, Software, Writing - Original
draft preparation. 
Qiuhao Wang: Investigation, Model Deployment.
Enze Zhou: Signal Acquisition, Android Development. 
Ling Fu: Supervision, Writing Guidance. 
Qian Liu: Funding Support, Investigation.
Gang Zheng: Hardware Design.

\textbf{Open Access} This article is licensed under a Creative Commons
Attribution 4.0 International License, which permits use, sharing, adaptation, distribution and reproduction in any medium or format, as
long as you give appropriate credit to the original author(s) and the
source, provide a link to the Creative Commons licence, and indicate if changes were made. The images or other third party material
in this article are included in the article’s Creative Commons licence,
unless indicated otherwise in a credit line to the material. If material
is not included in the article’s Creative Commons licence and your
intended use is not permitted by statutory regulation or exceeds the
permitted use, you will need to obtain permission directly from the copyright holder. To view a copy of this licence, visit
\url{http://creativecommons.org/licenses/by/4.0/}.


% \bibliographystyle{spmpsci}      % mathematics and physical sciences
\bibliographystyle{ieeetr}       % APS-like style for physics
\bibliography{cas-refs}   % name your BibTeX data base


\end{document}